\input amssym.def  

\documentstyle{article}

\parskip=\smallskipamount
\let\em\sl

\newcommand{\sect}{\setcounter{equation}{0}\section}  
\newcommand{\sss}[1]{{\scriptscriptstyle #1}}
\newcommand{\half}{{\textstyle{1\over2}}}
\newcommand{\fourth}{{\textstyle{1\over4}}}
\newcommand{\real}{\mathrm{I\kern-.2em R}}
\newcommand{\e}{\mathrm{e}}
\newcommand{\im}{\mathrm{i}}
\newcommand{\tr}{\mathop{\mathrm{tr}}\nolimits}

\mathchardef\hodge="0203
\newcommand{\Lie}{\mathord{\hbox{\pounds}}}
\undefine\hbar
\newsymbol\hbar 207E

\newcommand{\bndout}{{{\cal T}_{\mathrm{outer}}}}
\newcommand{\initial}{{{\mit\Sigma}_{\mathrm{initial}}}}
\newcommand{\final}{{{\mit\Sigma}_{\mathrm{final}}}}
\newcommand{\bnd}{{\cal T}}
\newcommand{\ini}{{\mit\Sigma}}
\newcommand{\bouter}{{{\cal B}_{\mathrm{outer}}}}
\newcommand{\binner}{{{\cal B}_{\mathrm{inner}}}}
\newcommand{\group}{\frak G}

\newcommand{\hb}{\hbar}
\newcommand{\ib}{{\bar\imath}}
\newcommand{\jb}{{\bar\jmath}}
\newcommand{\fa}{\frak a}
\newcommand{\fb}{\frak b}
\newcommand{\fc}{\frak c}
\newcommand{\fd}{\frak d}

\newcommand{\metric}{g}
\newcommand{\ff}{\frak f}
\newcommand{\Kilmet}{\frak g}
\newcommand{\normbnd}{n}
\newcommand{\fffbnd}{\gamma}
\newcommand{\sffbnd}{{\mit\Theta}}
\newcommand{\normini}{u}
\newcommand{\enormini}{{\bar\normini}}
\newcommand{\fffini}{h}
\newcommand{\sffini}{K}
\newcommand{\fffbndini}{\sigma}
\newcommand{\sffbndini}{k}

\newcommand{\projbnd}{{\hat\fffbnd}}
\newcommand{\projini}{{\hat\fffini}}
\newcommand{\projbndini}{{\hat\fffbndini}}

\renewcommand{\time}{t}
\newcommand{\lapse}{N}
\newcommand{\shift}{N}
\newcommand{\elapse}{{\bar\lapse}}
\newcommand{\eshift}{{\bar\shift}}
\newcommand{\accini}{a}
\newcommand{\radius}{r}
\newcommand{\elevate}{M}

\newcommand{\volman}{\sqrt{-\metric}\,}
\newcommand{\volbnd}{\sqrt{-\fffbnd}\,}
\newcommand{\volini}{\sqrt{\fffini}\,}
\newcommand{\volbndini}{\sqrt{\fffbndini}\,}
\newcommand{\evolman}{\sqrt{\metric}\,}
\newcommand{\evolbnd}{\sqrt{\fffbnd}\,}
\newcommand{\intman}{\int_{\cal M}d^4\!x\,}
\newcommand{\intbnd}{\int_\bnd d^3\!x\,}
\newcommand{\intini}{\int_\ini d^3\!x\,}
\newcommand{\intbndini}{\int_{\partial\ini} d^2\!x\,}

\newcommand{\del}{\nabla}
\newcommand{\delini}{\raise2pt\hbox{$\mathchar"0235$}}
\newcommand{\delbnd}{\mathchar"0234 }
\newcommand{\covdel}{\frak D}
\newcommand{\covdelini}{\frak d}
\newcommand{\var}{\delta}

\newcommand{\Ricci}{R}
\newcommand{\Einstein}{G}
\newcommand{\conn}{\frak A}
\newcommand{\fV}{\frak V}
\newcommand{\fW}{\frak W}
\newcommand{\Faraday}{\frak F}
\newcommand{\Maxwell}{(\hodge\frak F)}
\newcommand{\elec}{\frak E}
\newcommand{\magn}{\frak B}
\newcommand{\ffpot}{A}
\newcommand{\fffield}{{\mit\Lambda}}
\newcommand{\ffelec}{E}
\newcommand{\tfpot}{A}
\newcommand{\tffield}{H}
\newcommand{\tfelec}{E}
\newcommand{\tfmagn}{B}

\newcommand{\GFaction}{S_\sss{\mathrm{GF}}}
\newcommand{\totalaction}{S_{\mathrm{total}}}
\newcommand{\microaction}{S_{\mathrm{micro}}}
\newcommand{\grandaction}{S_{\mathrm{grand}}}
\newcommand{\eaction}{I}
\newcommand{\asteaction}{I_\ast}
\newcommand{\microeaction}{I_{\mathrm{micro}}}
\newcommand{\grandeaction}{I_{\mathrm{grand}}}
\newcommand{\REFaction}{S_\sss0}
\newcommand{\TOThamcon}{{\cal H}}
\newcommand{\TOTmomcon}{{\cal H}}
\newcommand{\gausscon}{{\cal G}}
\newcommand{\EHDaction}{S_\sss{\mathrm{EHD}}}
\newcommand{\HEHD}{{\Bbb H}_\sss{\mathrm{EHD}}}
\newcommand{\EHDham}{H_\sss{\mathrm{EHD}}}
\newcommand{\EHDlgr}{L_\sss{\mathrm{EHD}}}
\newcommand{\EHDhamcon}{{\cal H}_\sss{\mathrm{EHD}}}
\newcommand{\EHDmomcon}{({\cal H}_\sss{\mathrm{EHD}})}
\newcommand{\DYMaction}{S_\sss{\mathrm{DYM}}}
\newcommand{\HDYM}{{\Bbb H}_\sss{\mathrm{DYM}}}
\newcommand{\DYMham}{H_\sss{\mathrm{DYM}}}
\newcommand{\DYMlgr}{L_\sss{\mathrm{DYM}}}
\newcommand{\DYMhamcon}{{\cal H}_\sss{\mathrm{DYM}}}
\newcommand{\DYMmomcon}{({\cal H}_\sss{\mathrm{DYM}})}
\newcommand{\DYMgausscon}{({\cal G}_\sss{\mathrm{DYM}})}
\newcommand{\AYMaction}{S_\sss{\mathrm{AYM}}}
\newcommand{\HAYM}{{\Bbb H}_\sss{\mathrm{AYM}}}
\newcommand{\AYMham}{H_\sss{\mathrm{AYM}}}
\newcommand{\AYMlgr}{L_\sss{\mathrm{AYM}}}
\newcommand{\AYMhamcon}{{\cal H}_\sss{\mathrm{AYM}}}
\newcommand{\AYMmomcon}{({\cal H}_\sss{\mathrm{AYM}})}
\newcommand{\AYMgausscon}{({\cal G}_\sss{\mathrm{AYM}})}
\newcommand{\DFFaction}{S_\sss{\mathrm{DFF}}}
\newcommand{\HDFF}{{\Bbb H}_\sss{\mathrm{DFF}}}
\newcommand{\DFFham}{H_\sss{\mathrm{DFF}}}
\newcommand{\DFFlgr}{L_\sss{\mathrm{DFF}}}
\newcommand{\DFFhamcon}{{\cal H}_\sss{\mathrm{DFF}}}
\newcommand{\DFFgausscon}{({\cal G}_\sss{\mathrm{DFF}})}
\newcommand{\DTFaction}{S_\sss{\mathrm{DTF}}}
\newcommand{\HDTF}{{\Bbb H}_\sss{\mathrm{DTF}}}
\newcommand{\DTFham}{H_\sss{\mathrm{DTF}}}
\newcommand{\DTFlgr}{L_\sss{\mathrm{DTF}}}
\newcommand{\DTFhamcon}{{\cal H}_\sss{\mathrm{DTF}}}
\newcommand{\DTFmomcon}{({\cal H}_\sss{\mathrm{DTF}})}
\newcommand{\DTFgausscon}{({\cal G}_\sss{\mathrm{DTF}})}

\newcommand{\EOMgeo}{({\mit\Xi}_{\mathrm{geom}})}
\newcommand{\EOMdil}{{\mit\Xi}_{\mathrm{dil}}}
\newcommand{\EOMYM}{\frak X}
\newcommand{\EOMaxion}{{\mit\Xi}_{\mathrm{axi}}}
\newcommand{\EOMAYM}{\frak Y}
\newcommand{\EOMFF}{({\mit\Xi}_\sss{\mathrm{DFF}})}
\newcommand{\EOMTF}{({\mit\Xi}_\sss{\mathrm{DTF}})}
\newcommand{\Stress}{T}
\newcommand{\YMstress}{(T_\sss{\mathrm{DYM}})}
\newcommand{\YMdilsrc}{{\mit\Upsilon}_\sss{\mathrm{DYM}}}
\newcommand{\YMSRC}{\frak J}
\newcommand{\AYMstress}{(T_\sss{\mathrm{AYM}})}
\newcommand{\FFstress}{(T_\sss{\mathrm{DFF}})}
\newcommand{\FFdilsrc}{{\mit\Upsilon}_\sss{\mathrm{DFF}}}
\newcommand{\TFstress}{(T_\sss{\mathrm{DTF}})}
\newcommand{\TFdilsrc}{{\mit\Upsilon}_\sss{\mathrm{DTF}}}
\newcommand{\tfsrc}{J}

\newcommand{\SMDTOT}{{\cal J}}
\newcommand{\SCD}{{\cal Q}}
\newcommand{\SCUR}{{\cal I}}
\newcommand{\mombnd}{\pi}
\newcommand{\momini}{p}
\newcommand{\ssem}{\tau}
\newcommand{\SED}{{\cal E}}
\newcommand{\SMD}{({\cal J}_\sss{\mathrm{EHD}})}
\newcommand{\SSD}{{\cal S}}
\newcommand{\SEDREF}{\SED_\sss0}
\newcommand{\SMDREF}{({\cal J}_\sss{\mathrm{EHD,0}})}
\newcommand{\SSDREF}{(\SSD_\sss0)}
\newcommand{\SDD}{{\cal Y}}
\newcommand{\SDDREF}{\SDD_\sss0}

\newcommand{\shapebndini}{\varsigma}
\newcommand{\VOL}{{\cal V}}

\newcommand{\Pdil}{P_{\mathrm{dil}}}
\newcommand{\Pidil}{{\mit\Pi}_{\mathrm{dil}}}
\newcommand{\PYM}{(P_\sss{\mathrm{DYM}})}
\newcommand{\PiYM}{({\mit\Pi}_\sss{\mathrm{DYM}})}
\newcommand{\SYMCD}{({\cal Q}_\sss{\mathrm{DYM}})}
\newcommand{\SYMMD}{({\cal J}_\sss{\mathrm{DYM}})}
\newcommand{\SYMCUR}{\frak I}
\newcommand{\PAYM}{(P_\sss{\mathrm{AYM}})}
\newcommand{\PiAYM}{({\mit\Pi}_\sss{\mathrm{AYM}})}
\newcommand{\Paxion}{P_{\mathrm{axi}}}
\newcommand{\Piaxion}{{\mit\Pi}_{\mathrm{axi}}}
\newcommand{\SAYMCD}{({\cal Q}_\sss{\mathrm{AYM}})}
\newcommand{\SAYMMD}{({\cal J}_\sss{\mathrm{AYM}})}
\newcommand{\SAYMCUR}{\frak K}
\newcommand{\SAD}{{\cal A}}
\newcommand{\PFF}{(P_\sss{\mathrm{DFF}})}
\newcommand{\PiFF}{({\mit\Pi}_\sss{\mathrm{DFF}})}
\newcommand{\SFFCD}{({\cal Q}_\sss{\mathrm{DFF}})}
\newcommand{\PTF}{(P_\sss{\mathrm{DTF}})}
\newcommand{\PiTF}{({\mit\Pi}_\sss{\mathrm{DTF}})}
\newcommand{\STFCD}{({\cal Q}_\sss{\mathrm{DTF}})}
\newcommand{\STFMD}{({\cal J}_\sss{\mathrm{DTF}})}
\newcommand{\STFCUR}{({\cal I}_\sss{\mathrm{DTF}})}

\newcommand{\energy}{\Bbb E}
\newsymbol\EHDcharge 207C
\newcommand{\mass}{\Bbb M}
\newcommand{\anglmom}{\Bbb J}
\newcommand{\DYMcharge}{\Bbb Q_\sss{\mathrm{DYM}}}

\newcommand{\DFFcharge}{\Bbb Q_\sss{\mathrm{DFF}}}
\newcommand{\DTFcharge}{\Bbb Q_\sss{\mathrm{DTF}}}

\newcommand{\dil}{{\mit\Psi}}
\newcommand{\ddil}{{\mathaccent"0017 \dil}}
\newcommand{\ldil}{\psi}
\newcommand{\axion}{\theta}
\newcommand{\daxion}{{\mathaccent"0017 \axion}}
\newcommand{\Killing}{\xi}
\newcommand{\Killazm}{\varphi}
\newcommand{\Killgauge}{\frak k}
\newcommand{\rot}{\omega}
\newcommand{\rectemp}{\beta}
\newcommand{\period}{\Delta\time}
\newcommand{\spress}{{\cal P}}
\newcommand{\sshape}{\lambda}

\newcommand{\dens}{\varrho}
\newcommand{\ehgrav}{\kappa_\sss{\mathrm{H}}}
\newcommand{\entropy}{\Bbb S}
\newcommand{\feh}{f_\sss{\mathrm{EH}}(\dil)}
\newcommand{\fke}{f_\sss{\mathrm{KE}}(\dil)}
\newcommand{\fpe}{f_\sss{\mathrm{PE}}(\dil)}
\newcommand{\fym}{f_\sss{\mathrm{YM}}(\dil)}
\newcommand{\fff}{f_\sss{\mathrm{FF}}(\dil)}
\newcommand{\ftf}{f_\sss{\mathrm{TF}}(\dil)}
\newcommand{\fgf}{f_\sss{\mathrm{GF}}(\dil)}
\newcommand{\thym}{\vartheta_\sss{\mathrm{YM}}(\axion)}
\newcommand{\thke}{\vartheta_\sss{\mathrm{KE}}(\axion)}
\newcommand{\thpe}{\vartheta_\sss{\mathrm{PE}}(\axion)}

\begin{document}

\title{Quasilocal Thermodynamics of Dilaton Gravity\\
  coupled to Gauge Fields}
\author{Jolien D. E. Creighton%
  \thanks{e-mail: {\tt jolien@avatar.uwaterloo.ca}---on leave from
  Dept.~of Physics, University of Waterloo, Waterloo, Ontario, Canada}
  \ and Robert B. Mann%
  \thanks{e-mail: {\tt rbm20{@}amtp.cam.ac.uk}---on leave from
  Dept.~of Physics, University of Waterloo, Waterloo, Ontario, Canada}\\[1pc]
  {\em Department of Applied Mathematics and Theoretical Physics,}\\
  {\em Silver Street, Cambridge, England.  CB3\ 9EW}}
\date{19 April, 1995}
\maketitle

\begin{abstract}\noindent
We consider an Einstein-Hilbert-Dilaton action for gravity
coupled to various types of Abelian and non-Abelian gauge fields
in a spatially finite system.
These include Yang-Mills fields and Abelian gauge fields with
three and four-form field strengths.
We obtain various quasilocal quantities associated with these fields,
including their energy and angular momentum, and
develop methods for calculating conserved charges when a solution possesses
sufficient symmetry. For stationary black holes, we find an expression
for the entropy from the micro-canonical form of the action.  We also find
a form of the first law of black hole thermodynamics for black holes
with the gauge fields of the type considered here.
\end{abstract}

\vfill\pagebreak

\sect{Introduction}

The relationship between the Euclidean-action formulation of quantum
gravity and the thermodynamics of the gravitational field has been a
subject of increasing interest in recent years.  Fundamental
connections between the partition function of the grand canonical
ensemble and the Euclidean-action path integral were first pointed out
by Gibbons and Hawking \cite{GibH}, who argued that the Euclidean
gravitational action is equal to the grand canonical free energy times
the reciprocal of the temperature associated with a black hole (or
cosmological) event horizon \cite{BekH}.

More recently Brown and York
have extended this work by considering the formulation of the
partition function for gravitating systems of finite spatial extent
\cite{BYb,BYc}. Virtually all systems with which we have any experience
have a finite spatial boundary; indeed,
one of the central concepts in thermodynamics is that of a system and
a reservoir that are separated by a partition.
A physical realization of these concepts is needed in order to apply
thermodynamics in a sensible way.%
\footnote{Another central concept of thermodynamics is, of course,
thermodynamic
equilibrium.  The realization of this is the stationarity of a system,
where we say a system is stationary if there exists a time-like (Killing)
vector field for which the Lie derivative of all the fields considered vanish,
and the boundary of the system
is chosen to contain the orbits of these vectors.
We shall assume that our systems are stationary.}
When we use thermodynamics to describe self-gravitating systems (including
objects such as black holes), we must divide our space-time into a region
which contains the system (the black hole) and the remainder
of the space-time which can be treated as the reservoir.  The traditional
practice is to study black hole thermodynamics at space-like infinity,
assuming reasonable asymptotic conditions without any boundary.
However, this approach has a number of
deficiencies.  First, it requires that a spacetime
which is a solution to the field equations (and often the accompanying matter
matter fields) possess appropriate
asymptotic behaviour, typically asymptotic flatness. However asymptotic
flatness is never satisfied in reality, and is not always an
appropriate theoretical idealization: many black hole spacetimes
exist that are solutions to the (dilatonic) gravitational field equations that
do not possess asymptotic flatness. Some have been found recently that
are not even asymptotic to  de\thinspace Sitter or anti-de\thinspace Sitter
spacetime \cite{KJR}. Furthermore, in the study of pair creation of
black holes one is forced to consider non-asymptotically flat
spacetimes with an acceleration horizon \cite{gwg}, necessitating a
more careful consideration of boundary terms in the formulation of
the Hamiltonian \cite{HorH}.
Second, construction of a partition function (which is central in the study
of statistical mechanics) requires the stability of the system, which
is only realized when a finite size is imposed \cite{BYc}, a point
also noted by Hayward and Wong \cite{Hay}. For example,
the heat capacity for a Schwarzschild black hole is
negative~\cite{HawIs} if one fixes the temperature at
infinity, and the formal expression for the partition
function is not logically consistent~\cite{Waldin}. However if
the temperature is fixed at a finite
spatial boundary, there is
no inconsistency in the black hole partition function
and the heat capacity is positive ~\cite{York1}. This approach may
also be extended to black holes in anti de\thinspace Sitter spacetimes
\cite{BCM}, where analogous results are obtained.
Finally, it seems to us that, on physical grounds,
one should be able to define thermodynamics
appropriate to observers who are at a finite distance from the black hole.

It is important, therefore, to construct thermodynamic quantities appropriate
to observers at the (finite) boundary of a system.  The quasilocal
formalism, which has been developed extensively by Brown and York \cite{BYa}
provides
us with a means.  This formalism is based on a Hamilton-Jacobi principle
wherein the boundary terms of the action functional for a compact region
give rise to the quasilocal quantities such as energy and angular momentum.
Brown and York \cite{BYb} have also developed a paradigm for
understanding the relationship between the classical mechanics, the
statistical mechanics, and the thermodynamics of gravitating systems, based
on the boundary conditions of the system.  Here, the fields that are held
fixed on the boundary in deriving the classical mechanics equations of
motion from an action principle determine the statistical ensemble of the
corresponding statistical mechanics, and thus the type of thermodynamic
partition that must be imposed.  There is a parallelism between the Legendre
transformation---a canonical transformation---of the boundary terms of the
action, and the Laplace transformation---which changes the type of
ensemble---of the path integral for the statistical mechanics.
This parallelism
may be used to identify a ``micro-canonical'' action that
has boundary terms appropriate to a micro-canonical statistical
ensemble \cite{BYb}, and from it
an expression for the entropy of a thermodynamic
system may be obtained.

In this paper we extend this formalism to include the most
general action of gauge fields coupled to dilaton gravity
that has at most two derivatives in any term.
This class of actions includes the low-energy limit to string theories
\cite{string}
and is also of interest as an empirical foil for testing general
relativity \cite{Ni,Damour}.
We consider non-Abelian gauge theories coupled to
gravity, the dilaton, and an axion field, as well as
Abelian two-form and three-form
gauge potentials with three-form and four-form field strengths respectively,
also coupled to a dilation.  (We note that recent research in
non-Abelian gauge fields coupled to gravity
has led to new black hole solutions that refute much of the
folklore about black holes, such as the ``no hair'' conjecture
\cite{Wein}).
We will find that a form of the ``first law'' of
black hole thermodynamics can be
reconstructed for the theory considered which is quite reminiscent of the
conventional one for the Einstein-Maxwell theory.  Although we
restrict ourselves  to four dimensional
space-times, higher dimensional generalizations of our work
are straightforward. For each sector of the
action, we obtain expressions for the quasilocal thermodynamic quantities
that will appear in the first law of thermodynamics.
In addition, we discuss the
r\^ole of conserved quantities associated with the gauge and gravitational
fields.  In general,
the presence of a conserved quantity depends on the type of solution.  Such
quantities always require the presence of some sort of symmetry
in the solution.
(Solutions that satisfy certain asymptotic conditions often also possess
an asymptotic symmetry that can be used to define conserved quantities.
Here, the solution must at least possess exact symmetries on some finite
boundary in order to admit conserved
quantities for finite-sized systems.)  We also give a brief summary of the
construction of a statistical mechanics for the class of theories considered,
and we obtain an expression for the entropy and a form of the first law
of thermodynamics for systems possessing an event horizon.
We note that a version of the first law of black hole thermodynamics for
a finite system has been obtained for vacuum general relativity as well as
vacuum Einstein-Maxwell \cite{Hayward}; we
recover these results in the case of an Abelian gauge theory with vanishing
dilaton and axion fields.

Closely related to the work of Brown and York, as well as to our present
work, is that of Wald's Noether charge formalism \cite{Wald}.
Wald obtains an expression
for the entropy of a space-time solution of a very general class of theories
that are required to be derivable via an action principle from a Lagrangian
density that is covariant under diffeomorphisms.  Here, the entropy is
identified as the Noether charge associated with this covariance.  Recently,
Iyer and Wald \cite{IW}
have extended their treatment to systems with finite boundaries,
and they have shown how to formally recover some of the results of
Brown and York \cite{BYa}.
The present work can be viewed, therefore, as complementary to this technique.
We restrict our considerations to actions with at most two derivatives
in every term, and we explicity evaluate quantities which are only implicitly
defined in ref.~\cite{IW} for specific types of boundary conditions.
We justify the choices of boundary
conditions by appealing to the r\^ole they play in the connection of the
classical mechanics, the statistical mechanics, and the thermodynamics
discussed above.

Therefore, our agenda is the following:  In section two, we will study
the Einstein-Hilbert-Dilaton sector, which we will consider as our
gravitational theory in the absence of additional fields.  In section three,
we consider a Yang-Mills field that couples to both the metric and
the dilaton,  extending this to include axion couplings
in section four.  In section five, we turn to an Abelian gauge theory
involving a four-form field strength (also coupled to the dilaton), and briefly
consider its relation to a cosmological constant.  In section six we
complete our survey of matter fields with an Abelian three-form field
strength gauge theory coupled to the dilation.  Section seven is a review
of statistical mechanics based on path integral techniques.  Here we adopt
a ``general form'' for the gauge fields rather than treating each case
separately.  In section eight we apply these results to obtain an expression
for the entropy of a system containing a black hole and a form of the first
law of black hole thermodynamics.  Concluding remarks follow.

In this paper, we adopt the conventions of Wald \cite{WaldBook}.  Units are
chosen so that the speed of light, Newton's constant and the rationalized
Planck constant are all taken to be unity.
A summary of the notation of the manifolds, fields, and
related quantities considered in this paper is given in the appendix.


\sect{Einstein-Hilbert-Dilaton Sector}

The theory of gravity that we consider is one which is based on the
usual action of general relativity, but with the addition of a scalar
function that couples to the curvature called the dilaton.  In
general, it is possible to redefine the metric via a conformal transformation
so that the dilaton does not couple to the new curvature.  However,
as we shall see, such a transformation affects physical quantities such
as the entropy associated with black holes, so we consider the more general
case.

As indicated, our theory of gravitation will be based on an action
principle, and is closely related to the usual Einstein-Hilbert action.
The gravitational field equations may be deduced using Hamilton's principle
when considering variations in the geometry.  It is now known that serious
restrictions must be placed on the boundary of the space-time region in
order that such a variational principle be well defined.  In particular,
both the variation of the induced metric and its derivatives must be held fixed
on the boundary.  Alternately, the action may be supplemented with additional
boundary terms such that we need only fix the induced metric on the boundary.
We take the latter approach; from these boundary terms
many useful quantities can be defined, as we will show.

In this section, we look at the gravitational sector of the theory, and
we defer consideration of various types of matter to later sections.
Insofar as we restrict ourselves to dilatonic gravity, we will consider
a very general action.  The coupling to the curvature, the kinetic energy
of the dilaton, and the potential energy of the dilaton will all be
arbitrary functions of the dilaton (alone).  This practice will be continued
in the next section when we consider gauge fields with arbitrary couplings
to the dilaton.

We will restrict our considerations to a spatially finite
region of a four-dimensional
manifold, $\cal M$, which has a topology of
${\mit\Sigma}\times\real$ where $\mit\Sigma$ is a space-like hypersurface
and $\real$ is a real interval.  On this manifold, we define a metric,
$\metric_{\mu\nu}$, and its compatible derivative operator $\del_\mu$.
Objects with
Greek indices represent tensor quantities on the four dimensional manifold.
We will consider two parts
of the boundary of this manifold: the outer boundary
$\bndout=\partial{\mit\Sigma}\times{\cal I}$, and the initial space-like
hypersurface $\initial$.  The addition of an inner boundary and a final
space-like
hypersurface are trivial extensions of the present analysis that will
be important in following sections.  Here we shall refer to them simply
as $\bnd$ and $\ini$ respectively.

Associated with $\bnd$ is an outward directed space-like normal vector
$\normbnd^\mu$.  We can construct the first and second fundamental forms
$\fffbnd_{\mu\nu}=\metric_{\mu\nu}-\normbnd_\mu\normbnd_\nu$  and
$\sffbnd_{\mu\nu}=-\half\Lie_\normbnd\fffbnd_{\mu\nu}$.  These are to be
viewed as tensors on $\bnd$, and the indices $i$, $j$, etc., will denote
tensor quantities on $\bnd$.
However, we will use $\projbnd^\mu_i$, defined in an analogous manner
to $\fffbnd_{\mu\nu}$,
to denote the projection operator from $\cal M$ onto $\bnd$.
The derivative operator compatible with $\fffbnd_{ij}$ is $\delbnd_i$.

We can foliate $\cal M$ into space-like leaves ${\mit\Sigma}_t$ via the
parameter $t$ which is a co\"ordinate along $\real$; associate with it
the vector field $\time_\mu=\partial_\mu t$.  View the initial hypersurface as
one of these leaves, say, the $t=0$ one.  Little confusion arises from
dropping the $t$ index and considering an arbitrary leaf, $\ini$, in the
foliation.
There is a future directed time-like normal vector $\normini^\mu$
to $\ini$.
We may define the fundamental forms,
$\fffini_{\mu\nu}=\metric_{\mu\nu}+\normini_\mu\normini_\nu$ and
$\sffini_{\mu\nu}=-\half\Lie_\normini\fffini_{\mu\nu}$---these are viewed
as tensors on $\ini$, and such tensors will be shown with indices $\ib$, $\jb$,
etc.---and projection operator, $\projini^\mu_\ib$,
from $\cal M$ onto $\ini$.  We may also define the lapse and shift
of the foliation by $\lapse=-\time^\mu\normini_\mu$ and
$\shift^\ib=\projini_\mu^\ib\time^\mu$,
so that $\time^\mu=\lapse\normini^\mu+\shift^\mu$.
The derivative operator on $\ini$ compatible
with $\fffini_{\ib\jb}$ is given by $\delini_\ib$,
and this can be used to define the Ricci scalar on $\ini$: $\Ricci[\fffini]$.
It is also important to consider the boundary, $\partial\ini$
of $\ini$.  We require that, on the intersection of $\ini$ and $\bnd$,
$\normini^\mu\normbnd_\mu=0$.  The first and second fundamental forms on
$\partial\ini$ are
$\fffbndini_{\ib\jb}=\fffini_{\ib\jb}-\normbnd_\ib\normbnd_\jb$ and
$\sffbndini_{\ib\jb}=-\half\Lie_\normbnd\fffbndini_{\ib\jb}$.
The Latin indices $a$, $b$, etc., are used to denote tensors on
$\partial\ini$.  The projection operator
from $\bnd$ onto $\partial\ini$, $\projbndini^i_a$, is obtained from
$\fffbndini_{ij}=\fffbnd_{ij}+\normini_i\normini_j$.

A zero vorticity observer is one for whom the vorticity,
$\mbox{\boldmath$\varpi$}=\hodge(\mbox{\boldmath$v$}\wedge
\mbox{\boldmath$dv$})$, is zero,
where {\boldmath$v$} is the velocity of the observer.  It can be seen that
observers who are co-moving with a given foliation on the boundary
$\bnd$ are zero vorticity observers.  For such an observer, the velocity
is just the normal vector $\normini^\mu$, and the acceleration
of the normal vector is given by
$\accini^\mu=\normini^\nu\del_\nu\normini^\mu=\lapse^{-1}\fffini^{\mu\nu}
\del_\nu\lapse$.

A summary of the notation described above is given in Table 1 of the
appendix.

\subsection{Variation of the Action}

In accordance with the above considerations, we choose the action for
the gravitational sector to be:
\begin{eqnarray}
  \EHDaction &=& \intman\volman\bigl(\feh\Ricci[\metric]
  +\fke(\del\dil)^2+\fpe\bigr)\nonumber\\
  & &\qquad - 2\intbnd\volbnd\feh\tr(\sffbnd)\nonumber\\
  & &\qquad - 2\intini\volini\feh\tr(\sffini)
  \label{EHD Action}
\end{eqnarray}
where the dilaton field is given by $\dil$, and $\feh$, $\fke$, and
$\fpe$ are model-dependent functions of the dilaton.  They are arbitrary
insofar as they contain no fields other than the dilaton and no derivatives
of the dilaton.
Notice the presence of the two boundary terms: these are just what is needed
to make a well defined variational principle on the initial space-like
hypersurface and outer boundary as we shall see below.

Now we vary the geometry and dilaton field configurations.  The geometry
is varied subject to the (gauge) restriction that the leaves of foliation
remain orthogonal to the boundary.  (That is, we hold the boundary fixed
in that variations of the normal dual-vectors to the boundaries are
proportional
to the normal dual-vectors.)
For convenience, we vary the inverse metric, except on the boundaries where
we can write the variation in terms of the covariant induced metrics.
The induced variation in the action is given by:
\begin{eqnarray}
  \var\EHDaction &=& \intman\volman
  \bigl(\EOMgeo_{\mu\nu}\var\metric^{\mu\nu}+\EOMdil\var\dil\bigr)\nonumber\\
  & &\qquad + \intbnd(\mombnd^{ij}\var\fffbnd_{ij}+\Pidil\var\dil)\nonumber\\
  & &\qquad - \intini(\momini^{\ib\jb}\var\fffini_{\ib\jb}+\Pdil\var\dil).
  \label{Var EHD Action}
\end{eqnarray}
Here,
\begin{eqnarray}
  \EOMgeo_{\mu\nu}&=&\feh\Einstein_{\mu\nu}[\metric]
  +\half\fke\bigl(\metric_{\mu\nu}(\del\dil)^2\nonumber\\
  & &\qquad-2(\del_\mu\dil)(\del_\nu\dil)\bigr)+\half\metric_{\mu\nu}\fpe
  \label{EOM geometry}
\end{eqnarray}
(with
$\Einstein_{\mu\nu}[\metric]=\Ricci_{\mu\nu}[\metric]-\half\metric_{\mu\nu}
\Ricci[\metric]$) and
\begin{equation}
  \EOMdil={d\feh\over d\dil}\Ricci[\metric]
  +{d\fke\over d\dil}(\del\dil)^2-2\del^\mu\bigl(\fke\del_\mu\dil\bigr)
  +{d\fpe\over d\dil}
  \label{EOM dilaton}
\end{equation}
can be considered as equations in the following way: under variations of
the geometry and dilaton field that leave the geometry and dilaton
configurations on the boundary fixed, $\EOMgeo_{\mu\nu}$ and $\EOMdil$ will be
zero at an extremum of the action.  (In general, there will be contributions
to these equations of motion from any additional matter present.)

Alternately, we could consider variations of the geometry and dilaton
configurations in which the equations of motion are held fixed, but
the boundary geometry and dilaton configurations are varied.  Under
these variations, we see from equation (\ref{Var EHD Action}) that
we can define the momenta conjugate to the boundary geometry of
the initial and outer hypersurfaces as
$\mombnd^{ij}=(\var\EHDaction/\var\fffbnd_{ij})_\sss{\mathrm{CL}}$ and
$\momini^{\ib\jb}=-(\var\EHDaction/\var\fffini_{\ib\jb})_\sss{\mathrm{CL}}$
respectively.  (The subscripted `{\sc cl}'
emphasizes that the variations are amongst field configurations
satisfying the equations of motion.)  Explicitly, these are:
\begin{equation}
  \mombnd^{ij}=\volbnd\Bigl(\fffbnd^{ij}
  \bigl(\normbnd^\mu\partial_\mu\feh\bigr)
  +\feh\bigl(\sffbnd^{ij}-\fffbnd^{ij}\tr(\sffbnd)\bigr)\Bigr)
  \label{momentum bndgeom}
\end{equation}
and
\begin{equation}
  \momini^{\ib\jb}=-\volini\Bigl(\fffini^{\ib\jb}
  \bigl(\normini^\mu\partial_\mu\feh\bigr)
  +\feh\bigl(\sffini^{\ib\jb}-\fffini^{\ib\jb}\tr(\sffini)\bigr)\Bigr)
  \label{momentum inigeom}
\end{equation}

Similarly, there are momenta conjugate to the dilaton field on the
initial and outer hypersurfaces defined in an analogous way.  These
quantities are:
\begin{equation}
  \Pidil=\Bigl({\var\EHDaction\over\var\dil}\Bigr)_\sss{\mathrm{CL}}
  =\volbnd\Bigl(2\fke\normbnd^\mu\partial_\mu\dil
  -2{d\feh\over d\dil}\tr(\sffbnd)\Bigr)
  \label{momentum bnddil}
\end{equation}
and
\begin{equation}
  \Pdil=-\Bigl({\var\EHDaction\over\var\dil}\Bigr)_\sss{\mathrm{CL}}
  =-\volini\Bigl(2\fke\normini^\mu\partial_\mu\dil
  -2{d\feh\over d\dil}\tr(\sffini)\Bigr).
  \label{momentum inidil}
\end{equation}

\subsection{Quasilocal Quantities}

The $\bnd$ boundary momentum, $\mombnd^{ij}$, contains useful information
about the energy and momentum densities of the gravitational field
held within this boundary.  To identify these quantities, it is useful
to decompose the variation of the boundary metric, $\fffbnd_{ij}$, into
the various projections normal and onto the foliation.  Thus, we write:
\begin{equation}
  \var\fffbnd_{ij}=\projbndini^a_i\projbndini^b_j\var\fffbndini_{ab}
  -{2\over\lapse}\var\lapse\normini_i\normini_j
  -{2\over\lapse}\normini_{(i}\projbndini_{j)a}\var\shift^a.
  \label{decomp var bnd metric}
\end{equation}
The corresponding decomposition of the boundary momentum,
$\mombnd^{ij}$, conjugate to $\fffbnd_{ij}$ leads us to define the surface
energy density, the surface momentum density, and the surface stress
density as:
$$
  \SED=\volbndini\normini_i\normini_j\ssem^{ij},\qquad
  \SMD^a=-\volbndini\normini_i\projbndini_j^a\ssem^{ij},\quad\mathrm{and}\quad
  \SSD^{ab}=\volbndini\projbndini^a_i\projbndini^b_j\ssem^{ij}
$$
respectively, where
$\ssem^{ij}=2\mombnd^{ij}/\volbnd$ is the surface
stress-energy-momentum on $\bnd$.

We next use equation (\ref{momentum bndgeom}), and the relationship:
\begin{equation}
  \sffbnd_{ij}=\sffbndini_{ij}+\normini_i\normini_j(\normbnd^\mu\accini_\mu)
  +\normini_{(i}\projbndini^\ib_{j)}\normbnd^\jb\sffini_{\ib\jb}
  \label{decomp sff bnd}
\end{equation}
(recall that $\accini^\mu=\normini^\nu\del_\nu\normini^\mu$ is the acceleration
of the normal $\normini^\mu$) to obtain expressions for $\SED$, $\SMD^a$,
and $\SSD^{ab}$.  An immediate consequence of equation (\ref{decomp sff bnd})
is that $\tr(\sffbnd)=\tr(\sffbndini)-\normbnd^\mu\accini_\mu$.  Then,
we have:
\begin{equation}
  \SED=-2\volbndini\bigl(\normbnd^\mu\partial_\mu\feh
  -\feh\tr(\sffbndini)\bigr)
  -\SEDREF
  \label{sed}
\end{equation}
\begin{equation}
  \SMD^a=2\volbndini\feh\normbnd_\ib\projbndini_\jb^a\sffini^{\ib\jb}
  =-{2\volbndini\over\volini}\normbnd_\ib\projbndini^a_\jb\momini^{\ib\jb}
  -\SMDREF^a
  \label{smd}
\end{equation}
and
\begin{eqnarray}
  \SSD^{ab}&=&2\volbndini\biggl(\fffbndini^{ab}\normbnd^\mu\partial_\mu\feh
  \nonumber\\
  & &\qquad+\feh\Bigl(\sffbndini^{ab}-\fffbndini^{ab}
  \bigl(\tr(\sffbndini)-\normbnd^\mu\accini_\mu\bigr)\Bigr)\biggr)
  -\SSDREF^{ab}
  \label{ssd}
\end{eqnarray}
where $\SEDREF$, $\SMDREF^a$, and $\SSDREF$ are additional contributions that
arise in supplementing
the action $\EHDaction$ with an additional (reference) functional of the
boundary
fields, as we will discuss in the following sub-section.

We can construct a quantity, $\SDD$, from the momentum $\Pidil$ conjugate
to the dilaton configuration on $\bnd$: $\SDD=\lapse^{-1}\Pidil-\SDDREF$
where $\SDDREF$ is an arbitrary background contribution that will be discussed
below.  Thus we have:
\begin{equation}
  \SDD=\volbndini\Bigl(2\fke\normbnd^\mu\partial_\mu\dil-2{d\feh\over d\dil}\,
  \bigl(\tr(\sffbndini)-\normbnd^\mu\accini_\mu\bigr)\Bigr)-\SDDREF.
  \label{sdd}
\end{equation}
Note that $\SDD$ is also a scalar density on the two-surface $\partial\ini$.
Using the definitions in equations (\ref{sed}--\ref{ssd})
and (\ref{sdd}), we can write the $\bnd$ boundary terms in the variation of
the action as:
\begin{equation}
  \var\EHDaction\vert_\bnd=\intbnd\bigl(-\SED\var\lapse+\SMD_a\var\shift^a
  +\lapse(\half\SSD^{ab}\var\fffbndini_{ab}+\SDD\var\dil)\bigr).
  \label{bnd variation}
\end{equation}

It is worth noting that $\SED$, $\SMD_a$, $\fffbndini_{ab}$, and $\dil$
are all {\em extensive variables} because they can be constructed
out of the phase space variables, $(\momini^{\ib\jb},\fffini_{\ib\jb})$
and $(\Pdil,\dil)$ on $\ini$.
However the lapse and shift cannot be constructed
out of this information, and such quantities are called {\em intensive
variables}.  We see that, in equation (\ref{bnd variation}), the first two
terms of the integrand involve variations of intensive variables (with
extensive variables as coefficients), while the last three terms involve
variations of extensive variables (with coefficients that are intensive
variables due to the lapse function).  We will consider the implications
of this below.

Define the quasilocal energy to be the integral over the two-surface
$\partial\ini$ of the quasilocal energy density:
\begin{equation}
  \energy=\intbndini\SED.\label{quasilocal energy}
\end{equation}
This quantity has useful properties such as additivity \cite{BYa}, although it
is not necessarily positive definite.
The quasilocal energy is observer-dependent, i.e., even if we are given a
natural
choice of boundary $\bnd$, the above definition of the quasilocal energy
will still depend on the foliation; in particular, how the leaves $\ini$
intersect with $\bnd$.  However, we would expect that we should be able
to do better than this when the space-time is stationary.  The construction
of conserved charges resulting from space-times with symmetry will be
addressed below.

\subsection{Reference Space-time}

The complete gravitational action will involve the action $\EHDaction$ given in
equation (\ref{EHD Action}), plus some additional functional of the boundary
metric and dilaton configuration.  This additional piece, $\REFaction$,
does not contribute to the equations of motion because it contributes only
to the boundary.  It does, however, contribute to the momenta conjugate to
the gravitational and dilaton fields and thus to the quasilocal quantities
discussed above. It will be sufficient in the present analysis to consider
$\REFaction$ to be a functional on the boundary $\bnd$ alone.  Furthermore,
we will assume that it is a functional of the metric $\fffbnd_{ij}$, and
not $\bnd$ boundary dilaton configuration, but we will not consider the
general case in which it is also a functional of boundary configurations
of matter fields (though the inclusion of these is straight-forward).

The specification of the functional $\REFaction$ is akin to the specification
of some reference space-time.  Although there are restrictions on the types
of reference space-time allowed (which have to do with the embedding of the
two surfaces $\partial\ini$ in the reference spacetime), we will assume that
a natural reference space-time can be found.  Often, this is can be achieved
simply by setting constants of integration of a particular solution to some
special value that then specifies the reference.  Our restriction that the
functional $\REFaction$ not be a functional of matter fields
(other than the dilaton) is just a
specification of a vacuum reference space-time.

The primary advantage of the above interpretation of $\REFaction$ is that
it ensures that the action is a linear functional of lapse and shift and,
as we shall see, this means that the extensive variables defined above
continue to be extensive.  (This is the principal restriction that must
be placed on $\REFaction$.)  We shall thus write the additional piece in
the suggestive form:
\begin{equation}
  \REFaction=\intbnd\bigl(\lapse\SEDREF-\shift^a\SMDREF_a\bigr).
  \label{REF action}
\end{equation}
Here, $\SEDREF$ and $\SMDREF_a$ can be constructed out of the phase-space
variables alone (and are thus extensive), and can be identified by the
functional derivative of $\REFaction$ with respect to the lapse and the
(negative) shift respectively.  The surface-stress and dilaton density of the
reference spacetime can be obtained from the following relation:
\begin{equation}
\intbnd\lapse\bigl(\half\SSDREF^{ab}\var\fffbndini_{ab}+\SDDREF\var\dil\bigr)=
  -\intbnd\bigl(\lapse\var\SEDREF-\shift^a\var\SMDREF_a\bigr)
  \label{ssd REF}
\end{equation}
which allows one to determine $\SSDREF^{ab}$ and $\SDDREF$ when the form of
$\SEDREF$ and $\SMDREF_a$ are known in terms of $\fffbndini_{ab}$ and $\dil$.
The net result of the addition of $\REFaction$ to $\EHDaction$ is
the inclusion of the
extra terms in equations (\ref{sed}--\ref{ssd}) as well as
in the definition of $\SDD$.
Of course, one possible choice is always $\REFaction=0$ in which case
these terms would be absent.

\subsection{Conserved Charges}

Given a set of observers, all having histories on $\bnd$, we wish to
identify quantities related to the geometry of $\cal M$
that are the same for all the observers.  We will
require the presence of a Killing vector, $\Killing^i$, on $\bnd$.  Our
first step is to consider the equation of motion for the geometry.
We have been deferring a detailed discussion of matter to a later section;
however, here we will allow for the presence of matter so that the
equation of motion reads
\begin{equation}
  \EOMgeo_{\mu\nu}=\half\Stress_{\mu\nu}
\end{equation}
where $\Stress_{\mu\nu}$ is
the stress-energy-momentum tensor of the matter and $\EOMgeo_{\mu\nu}$ is
given by equation (\ref{EOM geometry}).  Computing
$\normbnd^\mu\projbnd^\nu_i\EOMgeo_{\mu\nu}$
with the aid of the Gauss-Codacci relationship
\begin{equation}
  \normbnd^\mu\projbnd^{i\nu}\Ricci_{\mu\nu}[\metric]=
  -\delbnd_j\bigl(\sffbnd^{ij}-\fffbnd^{ij}\tr(\sffbnd)\bigr)
  \label{Gauss Codacci 2}
\end{equation}
we obtain
\begin{equation}
  \volbnd\delbnd_j\ssem^{ij}=\Pidil\delbnd^i\dil
  -\normbnd^\mu\projbnd^{i\nu}\Stress_{\mu\nu}
  \label{div ssem}
\end{equation}
which shows that the surface stress-energy-momentum tensor is not
divergenceless: it has source terms arising from the presence of the
dilaton and the matter.

However, we can contract equation (\ref{div ssem}) with the Killing vector.
In addition to the usual requirements of a Killing vector, we require that
the dilaton is constant on orbits of the Killing vector.
That is, we assume that $\Killing^i$ satisfies both the usual Killing equation,
$\Lie_\Killing\fffbnd_{ij}=0$, for a Killing vector on $\bnd$, as well as
$\Lie_\Killing\dil=0$.  The left hand side of equation (\ref{div ssem})
becomes a total divergence due to the Killing equation and the symmetry
of $\ssem^{ij}$.  Integrating over $\bnd$, we find:
\begin{equation}
  -\int_{\bnd\cap\initial}^{\bnd\cap\final}d^2\!x\,
  \volbndini\Killing_i\normini_j\ssem^{ij}=
  \intbnd\volbnd\Killing^i\Stress_{i\mu}\normbnd^\mu.
  \label{cons geom charge}
\end{equation}
In the event that the right hand side vanishes for arbitrary $\final$,
equation (\ref{cons geom charge})
expresses a conservation law for a geometric charge,
\begin{equation}
  \EHDcharge[\Killing] = -\intbndini\volbndini\Killing_i\normini_j\ssem^{ij}.
  \label{geom charge}
\end{equation}
Contingent upon the type of matter present, there are many
reasons why this may be the case.   First, it may be that $\bnd$ is positioned
such
that there is little matter in its vicinity, so $\Stress_{\mu\nu}$ is
negligible.  Secondly, it may be the case that the particular projection
$\Killing^i\Stress_{i\mu}\normbnd^\mu$ vanishes.  Alternately, if
$\Stress_{\mu\nu}$
is conserved, that is $\del_\mu\Stress^{\mu\nu}=0$,
and if the Killing vector field on $\bnd$ can be promoted to a Killing
field over $\cal M$, (that is, if $\Killing^\mu$ is a Killing vector field
on $\cal M$, satisfying the additional requirements above, and $\bnd$
contains the orbits of these vectors) then we can re-write
equation (\ref{cons geom charge}) in the form:
\begin{equation}
  \EHDcharge(\bnd\cap\initial)-\EHDcharge(\bnd\cap\final)
  =\int_\initial^\final d^3\!x\,\volini\normini^\mu\Killing^\nu\Stress_{\mu\nu}
  \label{alt cons geom charge}
\end{equation}
and the integrand on the right hand side of this equation may vanish.

Suppose $\Killazm^i$ is a space-like azimuthal Killing vector.  Then, we
can define an angular momentum as $\anglmom=\EHDcharge[\Killazm]$.
If the surface $\partial\ini$ is taken so that it contains the orbits
of the Killing vector, then we can write:
\begin{equation}
  \anglmom=\intbndini\SMD_a\Killazm^a.
  \label{angular momentum}
\end{equation}

When $\Killing^i$ is time-like, we can define a mass as
$\mass=-\EHDcharge[\Killing]$.  If the spacetime is also static, that
is, $\Killing^i$ is surface forming, then we can choose a two-surface
$\partial\ini$ for which the Killing vector is proportional to the
time-like normal.  In this case, the mass can be written in the form:
\begin{equation}
  \mass=\intbndini\lapse\SED.
  \label{mass}
\end{equation}
Note that, in
general, the quasilocal energy will {\em not} agree with the conserved
mass of the space-time unless $\normini^i$ is a Killing vector on $\bnd$.
However, in asymptotically flat, static space-times for which $\bnd$ is taken
at
spatial infinity, a foliation in which $\normini^i$ approaches the time-like
Killing vector is usually adopted, and then the definitions of mass and
quasilocal energy given here will agree.

\subsection{Canonical Form of the Action}

We now turn to the canonical decomposition of the action of equation
(\ref{EHD Action}).  The Hamiltonian density of the Einstein-Hilbert-Dilaton
sector is defined as:
\begin{equation}
\EHDham=\momini^{\ib\jb}\Lie_\time\fffini_{\ib\jb}+\Pdil\Lie_\time\dil-\EHDlgr
  \label{ham dens}
\end{equation}
where $\EHDlgr$ is the Lagrangian density which, when integrated over
$\cal M$ yields the action of equation (\ref{EHD Action}).  To evaluate the
first term, we use the definition of the second fundamental form, and
the relationship between $\normini^\mu$ and $\time^\mu$.  We find that:
\begin{equation}
  \momini^{\ib\jb}\Lie_\time\fffini_{\ib\jb}=
  -2\lapse\momini^{\ib\jb}\sffini_{\ib\jb}
  -2\shift_\ib\delini_\jb\momini^{\ib\jb}
  +2\delini_\jb(\shift_\ib\momini^{\ib\jb}).
  \label{ham dens term 1}
\end{equation}
Similarly, the second term is
\begin{equation}
  \Pdil\Lie_\time\dil=\lapse\Pdil\ddil+\shift^\ib(\Pdil\delini_\ib\dil)
  \label{ham dens term 2}
\end{equation}
where the symbol $\ddil$ is shorthand for $\normini^\mu\partial_\mu\dil$.
The Lagrangian density can be canonically decomposed using the Gauss-Codacci
relationship:
\begin{equation}
  \Ricci[\metric]=\Ricci[\fffini]+\sffini^{\ib\jb}\sffini_{\ib\jb}
  -\bigl(\tr(\sffini)\bigr)^2
  -2\del_\mu\bigl(\normini^\mu\tr(\sffini)+\accini^\mu\bigr)
  \label{Gauss Codacci 1}
\end{equation}
as well as
\begin{equation}
  (\del\dil)^2=(\delini\dil)^2-\ddil^2.\label{dil ke decomp}
\end{equation}
We also use the relationship $\accini_\ib=\lapse^{-1}\delini_\ib\lapse$ which
is
appropriate for zero-vorticity observers.  The Hamiltonian can then be
written as:
\begin{eqnarray}
  \HEHD&=&\intini\EHDham\nonumber\\
  &=&\intini(\EHDhamcon\lapse+\EHDmomcon_\ib\shift^\ib)\nonumber\\
  & &\qquad+\intbndini(\SED\lapse-\SMD_a\shift^a)
  \label{EHD hamiltonian}
\end{eqnarray}
where,
\begin{eqnarray}
  \EHDhamcon&=&-2\momini^{\ib\jb}\sffini_{\ib\jb}+\Pdil\ddil
  \nonumber\\
  & &\qquad-\volini\biggl(\feh\Bigl(\Ricci[\fffini]
  +\sffini^{\ib\jb}\sffini_{\ib\jb}-\bigl(\tr(\sffini)\bigr)^2\Bigr)\nonumber\\
  & &\qquad+2\tr(\sffini){d\feh\over d\dil}\ddil-2\delini^2\feh\nonumber\\
  & &\qquad+\fke(\delini\dil)^2-\fke\ddil^2+\fpe\biggr)
  \label{EHD ham con}
\end{eqnarray}
is the Hamiltonian constraint and
\begin{equation}
  \EHDmomcon_\ib=-2\delini_\jb\momini_\ib^\jb+\Pdil\delini_\ib\dil
  \label{EHD mom con}
\end{equation}
is the momentum constraint.  When the constraint equations hold,
the Hamiltonian is purely a boundary term.

The action, equation (\ref{EHD Action}), can now be re-written in canonical
form:
\begin{eqnarray}
  \EHDaction&=&\intman\bigl(\momini^{\ib\jb}\Lie_\time\fffini_{\ib\jb}
  +\Pdil\Lie_\time\dil-\EHDham\bigr)\nonumber\\
  &=&\int dt\,\biggl(\,\intini\bigl(\momini^{\ib\jb}\Lie_\time\fffini_{\ib\jb}
  +\Pdil\Lie_\time\dil-\EHDhamcon\lapse-\EHDmomcon_\ib\shift^\ib\bigr)
  \nonumber\\
  & &\qquad\qquad+\intbndini(-\SED\lapse+\SMD_a\shift^a)\,\biggr).
  \label{EHD action canon}
\end{eqnarray}
This form of the action will be useful later in the study of the
thermodynamics of gravitational systems.


\sect{Yang-Mills Sector}

Here we consider non-Abelian gauge fields with an arbitrary gauge group
$\group$.
The case of electro-magnetism
is, of course, just a simplification of the general results of this
section when the gauge group is $U(1)$.  We shall refer to the internal degrees
of freedom
as colour degrees of freedom, and the associated gauge charges as colour
charges.

In what follows, quantities that possess colour are represented with
Fraktur characters; colour indices, when needed, will be given by lower
case Fraktur characters, and the adjoint representation will be assumed.
The gauge covariant derivative (both gauge covariant and covariant on
the manifold $\cal M$) is given by
$(\covdel_\mu)^\fa{}_\fb=\del_\mu\delta^\fa{}_\fb
+\ff^\fa{}_\fb{}_\fc\conn_\mu{}^\fc$
where $\conn_\mu{}^\fa$ is the connection and $\ff^\fa{}_\fb{}_\fc$ are the
structure constants of the group.  The curvature of this gauge
covariant derivative operator is the field tensor:
$\Faraday_{\mu\nu}{}^\fa=2\del_{[\mu}\conn_{\nu]}{}^\fa+\ff^\fa{}_\fb{}_\fc
\conn_\mu{}^\fb\conn_\nu{}^\fc$.  This field tensor is {\em co}variant under
gauge transformations of the form
$\conn_\mu{}^\fa[\frak x]=\conn_\mu{}^\fa[0]+(\covdel_\mu)^\fa{}_\fb\frak
x^\fb$.
We partially restrict the gauge freedom of the potential so that the
components in an orthonormal frame are finite everywhere in $\cal M$
(except, perhaps, at truly pathological points such as curvature
singularities).

On a space-like hypersurface, $\ini$, it is possible to decompose the
above quantities.  The electric field as seen by an observer
who is stationary with respect to the foliation is given by
$\elec_\ib{}^\fa=\projini_\ib^\mu\Faraday_{\mu\nu}{}^\fa\normini^\nu$.  The
magnetic field is given by
$\magn_\ib{}^\fa=-\half\projini_\ib^\mu\epsilon_{\mu\nu}{}^{\rho\sigma}
\Faraday_{\rho\sigma}{}^\fa\normini^\nu$.
The gauge covariant derivative that is compatible with
the metric $\fffini_{\ib\jb}$ is $(\covdelini_\ib)^\fa{}_\fb$.

\subsection{The Dilaton-Yang-Mills Action and Variations}

The action is
\begin{equation}
  \DYMaction=\intman\volman{1\over4}\fym(\Faraday^{\mu\nu}{}_\fa
  \Faraday_{\mu\nu}{}^\fa).
  \label{DYM action}
\end{equation}
The colour indices are raised and lowered by the Killing metric
$\Kilmet_\fa{}_\fb=\half\ff^\fc{}_\fd{}_\fa\ff^\fd{}_\fc{}_\fb$.  The function,
$\fym$, is a function of the dilaton alone, and contains no derivatives
of the dilaton.

Variation of the action of (\ref{DYM action}) yields source terms for
the Einstein-Hilbert field equations as well as for the dilaton field
equation.  In addition, variation with respect to the gauge potential
$\conn_\mu{}^\fa$ gives a source-free field equation for the gauge fields.
Note that the variation, on the boundary, of the potential is the same
as the variation of the potential projected onto the boundary.
The induced variation in the Dilaton-Yang-Mills sector of the
action is:
\begin{eqnarray}
  \var\DYMaction&=&
  \intman\volman\bigl(-\half\YMstress_{\mu\nu}\var\metric^{\mu\nu}
  -\half\YMdilsrc\var\dil+\EOMYM^\mu{}_\fa\var\conn_\mu{}^\fa\bigr)\nonumber\\
  & &\qquad+\intbnd\PiYM^i{}_\fa\var\conn_i{}^\fa
  -\intini\PYM^\ib{}_\fa\var\conn_\ib{}^\fa
  \label{var DYM action}
\end{eqnarray}
where
\begin{equation}
  \EOMYM^\nu{}_\fa=(\covdel_\mu)^\fb{}_\fa
  \bigl(\fym\Faraday^{\mu\nu}{}_\fb\bigr).
  \label{YM EOM}
\end{equation}
This yields the Dilaton-Yang-Mills vacuum equation of motion,
$\EOMYM^\nu{}_\fa=0$ when the boundary terms vanish provided that there
are no Yang-Mills source terms.
The Dilation-Yang-Mills stress energy and dilaton source are:
\begin{equation}
  \YMstress_{\mu\nu}=\fym(\Faraday_{\mu\alpha}{}^\fa\Faraday_\nu{}^\alpha{}_\fa
  -{\textstyle{1\over4}}\metric_{\mu\nu}
  \Faraday_{\alpha\beta}{}^\fa\Faraday^{\alpha\beta}{}_\fa)
  \label{YM stress}
\end{equation}
and
\begin{equation}
  \YMdilsrc={1\over2}\,{d\fym\over d\dil}\,\Faraday_{\mu\nu}{}^\fa
  \Faraday^{\mu\nu}{}_\fa
  \label{YM dil src}
\end{equation}
respectively.

In addition we have momenta conjugate to the variation of the gauge fields
on the boundaries $\bnd$ and $\ini$.  These are:
\begin{equation}
  \PiYM^i{}_\fa=\volbnd\fym\projbnd^i_\mu\Faraday^{\mu\nu}{}_\fa\normbnd_\nu
  \label{momentum bndYM}
\end{equation}
and
\begin{equation}
\PYM^\ib{}_\fa=-\volini\fym\projini^\ib_\mu\Faraday^{\mu\nu}{}_\fa\normini_\nu
  =-\volini\fym\elec^\ib{}_\fa
  \label{momentum iniYM}
\end{equation}
respectively.

On the boundary $\bnd$, we can decompose the variation of $\conn_i{}^\fa$ into
pieces normal and tangential to the foliation.
Let $\fV^\fa=\normini^i\conn_i{}^\fa$
and $\fW_a{}^\fa=\projbndini^i_a\conn_i{}^\fa$.  Then,
\begin{equation}
  \var\conn_i{}^\fa=\lapse^{-1}\normini_i\bigl(\var(\lapse\fV^\fa)
  -\fW_a{}^\fa\var\shift^a\bigr)+\projbndini^a_i\var\fW_a{}^\fa.
  \label{decomp var conn}
\end{equation}
A similar decomposition of $\PiYM^i_\fa$ leads us to define a
surface Yang-Mills charge density,
\begin{equation}
  \SYMCD_\fa=\volbndini\fym\normbnd_\ib\elec^\ib{}_\fa
  \label{symcd}
\end{equation}
and a surface Yang-Mills momentum density
\begin{equation}
  \SYMMD_a=\SYMCD_\fa\fW_a{}^\fa.
  \label{symmd}
\end{equation}
Also, define a surface Yang-Mills current,
\begin{equation}
  \SYMCUR_a{}^\fa=\volbndini\fym\epsilon_a{}^{\ib\jb}\normbnd_\ib
  \magn_\jb{}^\fa.
  \label{symcur}
\end{equation}
Then, the variation of $\DYMaction$ on $\bnd$ is given by:
\begin{equation}
  \var\DYMaction\vert_\bnd=\intbnd\bigl(\SYMMD_a\var\shift^a
  -\SYMCD_\fa\var(\lapse\fV^\fa)
  +\lapse\SYMCUR^a{}_\fa\var\fW_a{}^\fa\bigr).
  \label{YM bnd variation}
\end{equation}
We will consider the interpretation of the surface charge density in
the following.

\subsection{Conserved Yang-Mills Charges}

In the presence of a source, $\YMSRC^\mu{}_\fa$, the equations of motion for
the
Yang-Mills field (with dilaton coupling) are:
\begin{equation}
  (\covdel_\mu)^\fb{}_\fa\bigl(\fym\Faraday^{\mu\nu}{}_\fb\bigr)
  =\YMSRC^\nu{}_\fa.
  \label{YM EOM with source}
\end{equation}
It can be seen that the source respects the identity
$(\covdel_\mu)^\fb{}_\fa\YMSRC^\mu{}_\fb=0$.
In an Abelian gauge theory (such as electro-magnetism) this
quantity is gauge invariant and can be used to define a conserved charge.
However, in a general Yang-Mills theory, the identity is gauge-covariant,
and the separation of the colour contained in the charge and the colour
contained in the field it produces depends on the gauge choice.

Yet, it is still possible to construct a conserved colour charge
if we require the solution to the field equations to have certain
properties \cite{Abb,Town}.
Suppose that the solution possesses a gauge Killing scalar,
$\Killgauge^\fa$, that is, a Lie-algebra-valued scalar field on $\cal M$
that is covariantly constant: $(\covdel_\mu)^\fa{}_\fb\Killgauge^\fb=0$.
Then the quantity
$\Killgauge^\fa\YMSRC^\mu{}_\fa$ is gauge invariant and  divergenceless:
$\del_\mu(\Killgauge^\fa\YMSRC^\mu{}_\fa)=0$.  We can then define a charge,
\begin{equation}
  \DYMcharge[\Killgauge]=-\intini\volini\normini_\mu\Killgauge^\fa
  \YMSRC^\mu{}_\fa
  \label{conserved YM charge}
\end{equation}
that is conserved provided that the source $\Killgauge^\fa\YMSRC^\mu{}_\fa$
vanishes in the vicinity of $\bnd$, that is, the charge of equation
(\ref{conserved YM charge}) has the same value regardless of the volume
$\ini$ chosen for the integration.

Recall now the equation of motion (\ref{YM EOM with source}).  Contracting
both sides with the gauge Killing vector, one can take $\Killgauge^\fa$ through
the gauge covariant derivative to produce a gauge scalar as its argument.
Further contracting both sides by $\normini_\mu$ and
recalling that $\normini_\mu$ is proportional to a gradient,
$\normini_\mu=-\lapse\partial_\mu\time$,
we can write
$\normini_\nu(\covdel_\mu)^\fb{}_\fa\bigl(\fym\Faraday^{\mu\nu}{}_\fb\bigr)$ as
$(\covdelini_\mu)^\fb{}_\fa\bigl(\fym\Faraday^{\mu\nu}{}_\fb\normini_\nu\bigr)$,
due to the
antisymmetry of the field tensor $\Faraday^{\mu\nu}{}_\fa$ in $\mu$ and $\nu$.
Integrating over the space-like hypersurface $\ini$, we obtain
\begin{equation}
  \DYMcharge[\Killgauge]=\intbndini\SYMCD_\fa\Killgauge^\fa.
  \label{alt conserved YM charge}
\end{equation}
This alternate expression for the conserved gauge charge gives us our
interpretation of $\SYMCD_\fa$ as a surface gauge charge density.

\subsection{Canonical Form of the Dilaton-Yang-Mills Action}

Finally, we turn to the task of writing the Dilaton-Yang-Mills action
of equation (\ref{DYM action}) in canonical form.  A straight-forward
calculation shows that
$\Lie_\time\conn_\ib{}^\fa=-\Faraday_{\ib\jb}{}^\fa\time^\jb
-(\covdelini_\ib)^\fa{}_\fb\conn_\time{}^\fb$
where $\conn_\time{}^\fa=\time^\mu\conn_\mu{}^\fa$.
Using the decomposition of $\time^\mu$ into $\normini^\mu$ and
$\shift^\mu$, we have $\conn_\time{}^\fa=\lapse\fV^\fa-\shift^a\fW_a{}^\fa$.
Then, we can show that
\begin{eqnarray}
\PYM^\ib{}_\fa\Lie_\time\conn_\ib{}^\fa&=&-\lapse\PYM^\ib{}_\fa\elec_\ib{}^\fa
  -\PYM^\ib{}_\fa\Faraday_{\ib\jb}{}^\fa\shift^\jb\nonumber\\
  & &\qquad+\conn_\time{}^\fa
  (\covdelini_\ib)^\fb{}_\fa\PYM^\ib{}_\fb-\delini_\ib\bigl(\conn_\time{}^\fa
  \PYM^\ib{}_\fa\bigr).
  \label{YM ham dens term 1}
\end{eqnarray}
We also use the decomposition:
\begin{equation}
  \fourth\Faraday^{\mu\nu}{}_\fa\Faraday_{\mu\nu}{}^\fa=
  \half(\magn^\ib{}_\fa\magn_\ib{}^\fa-\elec^\ib{}_\fa\elec_\ib{}^\fa).
  \label{YM lgr decomp}
\end{equation}
The Hamiltonian density is given by
$\DYMham=\PYM^\ib{}_\fa\Lie_\time\conn_\ib{}^\fa-\DYMlgr$ where $\DYMlgr$
is the Lagrangian density (the integrand of equation (\ref{DYM action})).
The Hamiltonian is, then,
\begin{eqnarray}
  \HDYM&=&\intini\DYMham\nonumber\\
  &=&\intini\bigl(\DYMhamcon\lapse+\DYMmomcon_\ib\shift^\ib
  +\DYMgausscon_\fa\conn_\time{}^\fa\bigr)\nonumber\\
  & &\qquad+\intbndini\bigl(\SYMCD_\fa\fV^\fa\lapse-\SYMMD_a\shift^a\bigr).
  \label{YM hamiltonian}
\end{eqnarray}
where
\begin{equation}
  \DYMhamcon=-\PYM^\ib{}_\fa\elec_\ib{}^\fa+\half\volini\fym(\magn^\ib{}_\fa
  \magn_\ib{}^\fa-\elec^\ib{}_\fa\elec_\ib{}^\fa)
  \label{DYM ham con}
\end{equation}
is the contribution to the Hamiltonian constraint from the Dilaton-Yang-Mills
sector,
\begin{equation}
  \DYMmomcon_\jb=-\PYM^\ib{}_\fa\Faraday_{\ib\jb}{}^\fa
  \label{DYM mom con}
\end{equation}
is the contribution to the momentum constraint from the Dilaton-Yang-Mills
sector, and
\begin{equation}
  \DYMgausscon_\fa=(\covdelini_\ib)^\fb{}_\fa\PYM^\ib{}_\fb
  \label{DYM gauss con}
\end{equation}
is the Gauss constraint for the Yang-Mills field.

The action in canonical form is simply:
\begin{eqnarray}
  \DYMaction&=&\intman\bigl(\PYM^\ib{}_\fa\Lie_\time\conn_\ib{}^\fa
  -\DYMham\bigr)\nonumber\\
  &=&\int dt\,\biggl(\intini\bigl(\PYM^\ib{}_\fa\Lie_\time\conn_\ib^\fa
  \nonumber\\
  & &\qquad\qquad-\DYMhamcon\lapse-\DYMmomcon_\ib\shift^\ib-\DYMgausscon_\fa
  \conn_\time{}^\fa\bigr)\nonumber\\
  & &\qquad\qquad+\intbndini\bigl(-\SYMCD_\fa\fV^\fa\lapse
  +\SYMMD_a\shift^a\bigr)\biggl).
  \label{DYM action canon}
\end{eqnarray}


\sect{Axion Sector}

In this section we consider the coupling of an axion field to the
Yang-Mills field strength.  Such couplings provide an interesting
counterexample \cite{Ni} to  Schiff's conjecture\footnote{Schiff's
conjecture states that any self-consistent theory of
gravity that obeys the weak equivalence principle necessarily obeys
the Einstein Equivalence principle} \cite{Schiff} and, in the case of
an Abelian gauge theory, imply interesting new tests of the
equivalence principle \cite{Car,Haug}.

Many of the derivations in this section are similar to those in
the previous section on the Yang-Mills
field.  Define the dual to the Yang-Mills field tensor by
$\Maxwell_{\mu\nu}{}^\fa=\half\epsilon_{\mu\nu}{}^{\rho\sigma}
\Faraday_{\rho\sigma}{}^\fa$.
Note that there is the identity:
$(\covdel_\mu)^\fb{}_\fa\Maxwell^{\mu\nu}{}_\fb=0$.

\subsection{The Axion-Yang-Mills Action and its Variation}

We take the action for the Axionic sector to be
\begin{equation}
  \AYMaction=\intman\volman\bigl(\fourth\thym\Maxwell^{\mu\nu}{}_\fa
  \Faraday_{\mu\nu}{}^\fa+\thke(\del\axion)^2+\thpe\bigr)
  \label{AYM action}
\end{equation}
where $\axion$ is the axion field and $\thym$, $\thke$, and $\thpe$ are
functions of the axion (but not to its derivatives) that are the couplings
to the Yang-Mills fields, the Kinetic energy, and the Potential energy
respectively.

Varying the action with respect to the geometry, the gauge field, and the
axion yields
\begin{eqnarray}
  \var\AYMaction&=&\intman\volman\bigl(-\half\AYMstress_{\mu\nu}\var
  \metric^{\mu\nu}+\EOMAYM^\mu{}_\fa\var\conn_\mu{}^\fa
  +\EOMaxion\var\axion\bigr)\nonumber\\
  & &\qquad+\intbnd\bigl(\PiAYM^i{}_\fa\var\conn_i{}^\fa
  +\Piaxion\var\axion\bigr)\nonumber\\
  & &\qquad-\intini\bigl(\PAYM^\ib{}_\fa\var\conn_\ib{}^\fa
  +\Paxion\var\axion\bigr)
  \label{var AYM action}
\end{eqnarray}
where the stress-energy-momentum contribution from this sector is:
\begin{equation}
  \AYMstress_{\mu\nu}=-2\thke(\del_\mu\axion)(\del_\nu\axion)
  +\metric_{\mu\nu}\thke(\del\axion)^2+\metric_{\mu\nu}\thpe.
  \label{AYM stress}
\end{equation}
The equation of motion for the gauge field, when boundary variations
vanish, is $\EOMAYM^\mu{}_\fa=0$ (vacuum case) where:
\begin{equation}
  \EOMAYM^\mu{}_\fa=\Maxwell^{\mu\nu}{}_\fa\del_\nu\thym
  \label{EOM AYM}
\end{equation}
while the equation of motion for the axion itself is $\EOMaxion=0$ with:
\begin{equation}
  \EOMaxion={1\over4}{d\thym\over d\axion}\,\Maxwell^{\mu\nu}{}_\fa
  \Faraday_{\mu\nu}{}^\fa
  +{d\thke\over d\axion}\bigl((\del\axion)^2-2\del^2\axion\bigr)
  +{d\thpe\over d\axion}.
  \label{EOM axion}
\end{equation}
Similarly, there are momenta conjugate to the gauge field configurations
on $\bnd$ and $\ini$,
\begin{equation}
  \PiAYM^i{}_\fa=\volbnd\thym\normbnd_\mu\projbnd^i_\nu\Maxwell^{\mu\nu}{}_\fa
  \label{momentum AYM bnd}
\end{equation}
and
\begin{equation}
  \PAYM^\ib{}_\fa=-\volini\thym\normini_\mu\projini^\ib_\nu
  \Maxwell^{\mu\nu}{}_\fa
  \label{momentum AYM ini}
\end{equation}
respectively.  Also, there are momenta on $\bnd$ and $\ini$ conjugate to
the axionic field configurations:
\begin{equation}
  \Piaxion=2\volbnd\thke\normbnd^\mu\partial_\mu\axion\qquad\mathrm{and}\qquad
  \Paxion=-2\volini\thke\normini^\mu\partial_\mu\axion
  \label{momentum axion bnd ini}
\end{equation}
respectively.

Recall that the variation of the gauge field on the boundary $\bnd$
can be decomposed as in equation (\ref{decomp var conn}).  We perform
a similar decomposition of the momentum $\PiAYM^i{}_\fa$.  Define
a surface axion-Yang-Mills charge density,
\begin{equation}
  \SAYMCD_\fa=\volbndini\thym\normbnd_\ib\magn^\ib{}_\fa
  \label{saymcd}
\end{equation}
a surface axion-Yang-Mills momentum density,
\begin{equation}
  \SAYMMD_a=\SAYMCD_\fa\fW_a{}^\fa
  \label{saymmd}
\end{equation}
and a surface axion-Yang-Mills current,
\begin{equation}
  \SAYMCUR_a{}^\fa=\volbndini\thym\epsilon_a{}^{\ib\jb}\normbnd_\ib
  \elec_\jb{}^\fa.
  \label{saymcur}
\end{equation}
Also define the scalar density
\begin{equation}
  \SAD=N^{-1}\Piaxion.
  \label{sad}
\end{equation}
Then, the $\bnd$ portion
of the variation of $\AYMaction$ is decomposed into:
\begin{equation}
  \var\AYMaction\vert_\bnd=\intbnd\bigl(\SAYMMD_a\var\shift^a
  -\SAYMCD_\fa\var(\lapse\fV^\fa)+\lapse(\SAYMCUR^a{}_\fa\var\fW_a{}^\fa
  +\SAD\var\axion)\bigr).
  \label{AYM bnd variation}
\end{equation}
In analogy with the discussion of conserved charges in the previous
section, we can interpret the quantity $\SAYMCD_\fa$ to be some sort
of magnetic charge surface density that will yield a magnetic charge:
\begin{equation}
  \DFFcharge[\Killgauge]=\intbndini\SAYMCD_\fa\Killgauge^\fa
  \label{cons YM mag charge}
\end{equation}
when some gauge Killing scalar, $\Killgauge^\fa$, is present.

\subsection{Canonical Form of the Axion-Yang-Mills Action}

The Hamiltonian density of the Axion-Yang-Mills sector is given by
$\AYMham=\PAYM^\ib{}_\fa\Lie_\time\conn_\ib{}^\fa+\Paxion\Lie_\time\axion
-\AYMlgr$.  Denote by $\daxion$ the quantity $\normini^\mu\partial_\mu\axion$.
The first term in the Hamiltonian density looks just like equation
(\ref{YM ham dens term 1}) but with $\PAYM^\ib{}_\fa$ replacing
$\PYM^\ib{}_\fa$ everywhere.  The second term is just like equation
(\ref{ham dens term 2}), but here we replace $\Pdil$ and $\dil$ with
$\Paxion$ and $\axion$.  In addition, we can decompose the terms in
the Lagrangian density, $\AYMlgr$, which is just the integrand of
equation (\ref{AYM action}).  Note that:
\begin{equation}
  \fourth\thym\Maxwell^{\mu\nu}{}_\fa\Faraday_{\mu\nu}{}^\fa=
  \half\thym\magn^\ib{}_\fa\elec_\ib{}^\fa.
  \label{decomp AYM lgr}
\end{equation}
Furthermore, the kinetic term of the axion can be decomposed just
as the kinetic term of the dilaton was decomposed in equation
(\ref{dil ke decomp}).  Thus, the Hamiltonian is:
\begin{eqnarray}
  \HAYM&=&\intini\AYMham\nonumber\\
  &=&\intini\bigl(\AYMhamcon\lapse+\AYMmomcon_\ib\shift^\ib+\AYMgausscon_\fa
  \conn_t{}^\fa\bigr)\nonumber\\
  & &\qquad+\intini\bigl(\SAYMCD_\fa\fV^\fa\lapse-\SAYMMD_a\shift^a\bigr).
  \label{AYM hamiltonian}
\end{eqnarray}
where
\begin{eqnarray}
  \AYMhamcon&=&-\PAYM^\ib{}_\fa\elec_\ib{}^\fa+\Paxion\daxion
  -\volini\bigl(\half\thym\magn^\ib{}_\fa\elec_\ib{}^\fa\nonumber\\
  & &\qquad+\thke\bigl((\delini\axion)^2-\daxion^2\bigr)+\thpe\bigr)
  \label{AYM ham con}
\end{eqnarray}
is the Axion-Yang-Mills sector contribution to the Hamiltonian constraint,
\begin{equation}
  \AYMmomcon_\jb=-\PAYM^\ib{}_\fa\Faraday_{\ib\jb}{}^\fa
  +\Paxion\delini_\jb\axion
  \label{AYM mom con}
\end{equation}
is the Axion-Yang-Mills sector contribution to the momentum constraint, and
\begin{equation}
  \AYMgausscon_\fa=(\covdelini_\ib)^\fb{}_\fa\PAYM^\ib{}_\fb
  \label{AYM gauss con}
\end{equation}
is the Axion-Yang-Mills sector contribution to the Yang-Mills gauss
constraint.

The action of equation (\ref{AYM action}) may be written in canonical
form as
\begin{eqnarray}
  \AYMaction&=&\intman\bigl(\PAYM^\ib{}_\fa\Lie_\time\conn_\ib{}^\fa
  +\Paxion\Lie_\time\axion-\AYMham\bigr)\nonumber\\
  &=&\int dt\,\biggl(\intini\bigl(\PAYM^\ib{}_\fa\Lie_\time\conn_\ib{}^\fa
  +\Paxion\Lie_\time\axion\nonumber\\
  & &\qquad\qquad-\AYMhamcon\lapse-\AYMmomcon_\ib\shift^\ib
  -\AYMgausscon_\fa\conn_\time{}^\fa\bigr)\nonumber\\
  &
&\qquad\qquad+\intbnd\bigl(-\SAYMCD_\fa\fV^\fa\lapse+\SAYMMD_a\shift^a\bigr)
  \biggr).
  \label{AYM action canon}
\end{eqnarray}


\sect{Dilaton Four-Form Sector}

It is possible to treat a cosmological constant as a constant of motion
arising from a four-form field rather than as a fundamental constant.
Doing so allows one to consider space-times in which the cosmological
constant takes different values.  In this section, we will consider such
a four-form field, but here we will also allow possible couplings to the
dilaton.

The four-form field strength will be given by
$\fffield_{\mu\nu\rho\sigma}=4!\,\del_{[\mu}\ffpot_{\nu\rho\sigma]}$
where $\ffpot_{\lambda\mu\nu}$ is a three-form potential.  The field strength
is
invariant under gauge transformations of the potential of the form
$\ffpot_{\lambda\mu\nu}[\chi]=\ffpot_{\lambda\mu\nu}[0]+3!\,\del_{[\lambda}
\chi_{\mu\nu]}$ where $\chi_{\mu\nu}$ is an arbitrary two-form.
We partially restrict this gauge invariance in requiring the components
in an orthonormal basis be finite everywhere in $\cal M$.
It will be useful to introduce an `electric' field,
$\ffelec_{\hb\ib\jb}=\projini^\mu_\hb\projini^\nu_\ib\projini^\rho_\jb
\fffield_{\mu\nu\rho\sigma}\normini^\sigma$.  One can show that
$\fffield^{\mu\nu\rho\sigma}\fffield_{\mu\nu\rho\sigma}=-4\ffelec^{\hb\ib\jb}
\ffelec_{\hb\ib\jb}$.

\subsection{The Dilaton-Four-Form Action}

Let $\fff$ be a function of the dilaton (that contains no derivatives of
the dilaton) that couples to the four-form field as follows:
\begin{equation}
  \DFFaction=\intman\volman{1\over2\cdot4!}\,\fff\fffield^{\lambda\mu\nu\rho}
  \fffield_{\lambda\mu\nu\rho}.
  \label{DFF action}
\end{equation}
We could replace the four-form field strength $\fffield_{\lambda\mu\nu\rho}$
with its dual $\hodge\fffield$.  The result would be a cosmological constant
term with dilaton couplings.  We will not do so here as it is more useful
to consider variations of the three-form potential.  We will discuss the
relationship between the four-form field strength and the cosmological
constant below.

Under variations of the three-form potential, the geometry, and the dilaton
field configurations, the induced variation in the action of equation
(\ref{DFF action}) is:
\begin{eqnarray}
  \var\DFFaction&=&\intman\volman\bigl(-\half\FFstress_{\mu\nu}
  \var\metric^{\mu\nu}-\half\FFdilsrc\var\dil+\EOMFF^{\lambda\mu\nu}
  \var\ffpot_{\lambda\mu\nu}\bigr)\nonumber\\
  & &\qquad+\intbnd\PiFF^{ijk}\var\ffpot_{ijk}-\intini\PFF^{\hb\ib\jb}
  \ffpot_{\hb\ib\jb}.
  \label{var DFF action}
\end{eqnarray}
Note that the projection of the variation of the potential onto the
boundary elements is the same as the variation of the projection of
the potential onto the boundary elements.
Here, the stress-energy-momentum of the four-form field is given by:
\begin{equation}
  \FFstress_{\mu\nu}={1\over3!}\fff\Bigl({1\over8}\,\metric_{\mu\nu}
  \fffield^{\alpha\beta\gamma\delta}\fffield_{\alpha\beta\gamma\delta}
  -\fffield_\mu{}^{\alpha\beta\gamma}\fffield_{\nu\alpha\beta\gamma}\Bigr)
  \label{FF stress}
\end{equation}
and the dilaton source is:
\begin{equation}
  \FFdilsrc=-{1\over4!}\,{d\fff\over d\dil}\fffield^{\mu\nu\rho\sigma}
  \fffield_{\mu\nu\rho\sigma}.
  \label{FF dil src}
\end{equation}
The equations of motion for the four-form fields are:
\begin{equation}
  \EOMFF^{\lambda\mu\nu}=-\del_\alpha\bigl(\fff\fffield^{\alpha\lambda\mu\nu}
  \bigr).
  \label{EOM FF}
\end{equation}
In addition, the momentum conjugate to the three-form potential on the
boundary $\bnd$ is given by:
\begin{equation}
  \PiFF^{ijk}=\volbnd\fff\normbnd_\alpha\fffield^{\alpha\lambda\mu\nu}
  \projbnd^i_\lambda\projbnd^j_\mu\projbnd^k_\nu
  \label{momentum DFF bnd}
\end{equation}
while the momentum conjugate on the boundary $\ini$ is:
\begin{equation}
  \PFF^{\hb\ib\jb}=-\volini\fff\normini_\alpha\fffield^{\alpha\lambda\mu\nu}
  \projini^\hb_\lambda\projini^\ib_\mu\projini^\jb_\nu
  =\volini\fff\ffelec^{\hb\ib\jb}.
  \label{momentum DFF ini}
\end{equation}

Since $\ffpot_{ijk}$ is proportional to the volume element $\epsilon_{ijk}$
on $\bnd$, the quantity $V_{ij}=\ffpot_{ijk}\normini^k$ will be a tensor
on $\partial\ini$, that is, $V_{ij}=\projini^a_i\projini^b_j V_{ab}$.
With this in mind, we see that the variation of $\ffpot_{ijk}$ can be
decomposed as follows:
\begin{equation}
  \var\ffpot_{ijk}=-3\lapse^{-1}\var\lapse\normini_{[i}V_{jk]}
  -3\normini_{[i}\projbndini^a_j\projbndini^b_{k]}\var V_{ab}.
  \label{decomp var ffpot}
\end{equation}
With the definition
\begin{equation}
  \SFFCD^{ab}=3\volbndini\fff\normbnd_\ib\ffelec^{\ib ab}
  \label{sffcd}
\end{equation}
for the four-form surface charge density,
the component of the variation of the action $\DFFaction$ on the boundary
$\bnd$ can be written:
$$\var\DFFaction\vert_\bnd=-\intbnd\SFFCD^{ab}\var(\lapse V_{ab}).$$
As usual, $\SFFCD^{ab}$ is related to a surface charge density.

\subsection{The Conserved Charge of the Four-Form Field}

Recall the equation of motion for the four-form field strength given by
equation (\ref{EOM FF}).  In the presence of a source, $J^{\lambda\mu\nu}$,
we have:
\begin{equation}
  -\del_\alpha\bigl(\fff\fffield^{\alpha\lambda\mu\nu}\bigr)
  =J^{\lambda\mu\nu}
  \label{EOM FF src}
\end{equation}
and the identity $\del_\lambda J^{\lambda\mu\nu}$ follows.  Let the
quantity $\epsilon_{\mu\nu}$ be the volume element on the surface
$\partial\ini$.  Then, $\del_\lambda(J^{\lambda\mu\nu}\epsilon_{\mu\nu})=0$.
If there is no source present in the vicinity of $\bnd$, then we see that
\begin{equation}
  \DFFcharge[\epsilon]=\intini\volini\rho
  \label{cons FF charge}
\end{equation}
is a conserved charge of the four-form field: its value is independent of
the choice of leaf $\ini$.%
\footnote{That is, given a two-form, $\epsilon_{\mu\nu}$, the associated
charge is independent of the foliation.  However, the interpretation
of this two-form as the volume element of the boundary of the leaves
of the foliation is, of course, foliation dependent.}
Here, $\rho=-3\epsilon_{ab}J^{ab\mu}\normini_\mu$
is the charge density.

Using our definition of $\rho$, we can express it in terms of the four-form
field strength by means of the equation of motion, (\ref{EOM FF src}).  Because
the unit normal to $\ini$ is proportional to a gradient (recall
$\normini_\mu=-\lapse\partial_\mu\time$), it can be pulled inside the
derivative
operator due to the antisymmetry of the field strength; in doing so, the
derivative operator on $\cal M$ becomes the one on $\ini$.  We find that
$\volini\rho=3\volini\delini_\ib\bigl(\fff\ffelec^{\ib ab}\epsilon_{ab}\bigr)$
and thus,
\begin{equation}
  \DFFcharge[\epsilon]=\intbndini\SFFCD^{ab}\epsilon_{ab}.
  \label{alt cons FF charge}
\end{equation}
Equation (\ref{alt cons FF charge}) justifies our interpretation of
$\SFFCD^{ab}$ as a surface density of four-form charge.

\subsection{Canonical Decomposition of the Dilaton-Four-Form Action}

To obtain the Hamiltonian density,
$\DFFham=\PFF^{\hb\ib\jb}\Lie_\time\ffpot_{\hb\ib\jb}-\DFFlgr$,
we compute:
\begin{eqnarray}
  \PFF^{\hb\ib\jb}\Lie_\time\ffpot_{\hb\ib\jb}&=&\lapse\Bigl(-{1\over3!}\,
  \volini\fff\ffelec^{\hb\ib\jb}\ffelec_{\hb\ib\jb}\Bigr)\nonumber\\
  & &\qquad-3\ffpot_{\ib\jb\time}\delini_\hb\PFF^{\hb\ib\jb}
  +\delini_\hb\bigl(3\lapse V_{ab}\PFF^{\hb ab}\bigr).
  \label{DFFham term 1}
\end{eqnarray}
Here, $\ffpot_{\ib\jb\time}=\ffpot_{\ib\jb\mu}\time^\mu$.
We have already seen that the Lagrangian density is purely electric:
\begin{equation}
  \DFFlgr=-\lapse\volini{1\over2\cdot3!}\fff\ffelec^{\hb\ib\jb}
  \ffelec_{\hb\ib\jb}.
  \label{DFFham term 2}
\end{equation}
Thus, the Hamiltonian is:
\begin{eqnarray}
  \HDFF&=&\intini\DFFham\nonumber\\
  &=&\intini\bigl(\DFFhamcon\lapse+\DFFgausscon^{\ib\jb}\ffpot_{\ib\jb\time}
  \bigr)+\intbndini\SFFCD^{ab}\lapse V_{ab}
  \label{DFF hamiltonian}
\end{eqnarray}
where the four-form contribution to the Hamiltonian constraint is:
\begin{equation}
  \DFFhamcon=-\volini{1\over2\cdot3!}\fff\ffelec^{\hb\ib\jb}\ffelec_{\hb\ib\jb}
  \label{DFF ham con}
\end{equation}
and the four-form Gauss constraint is:
\begin{equation}
  \DFFgausscon^{\ib\jb}=-3\delini_\hb\PFF^{\hb\ib\jb}.
  \label{DFF gauss con}
\end{equation}

The action in canonical form is:
\begin{eqnarray}
  \DFFaction&=&\intman\bigl(\PFF^{\hb\ib\jb}\Lie_\time
  \ffpot_{\hb\ib\jb}-\DFFham\bigr)\nonumber\\
  &=&\int dt\biggl(\intini\bigl(\PFF^{\hb\ib\jb}\Lie_\time\ffpot_{\hb\ib\jb}
  -\DFFhamcon\lapse-\DFFgausscon^{\ib\jb}\ffpot_{\ib\jb\time}\bigr)\nonumber\\
  & &\qquad\qquad-\intbndini\SFFCD^{ab}\lapse V_{ab}.
  \label{var canon DFF action}
\end{eqnarray}

\subsection{Relation to a Cosmological Constant}

Recall the Lagrangian density
\begin{eqnarray}
  \DFFlgr&=&\volman(2\cdot4!)^{-1}\fff
  \fffield^{\mu\nu\rho\sigma}\fffield_{\mu\nu\rho\sigma}\nonumber\\
  &=&-\half\volman\fff(\hodge\fffield)^2.
  \label{DFF lgr}
\end{eqnarray}
In the latter form, it appears like a dilaton coupled
to a cosmological constant $\lambda=-\fourth(\hodge\fffield)^2$, except
that, where a cosmological constant is fixed in the theory, $\hodge\fffield$
is a scalar field.  The stress-energy-momentum tensor can be evaluated and
we find that $\FFstress_{\mu\nu}
=\half\fff(\hodge\fffield)^2\metric_{\mu\nu}$ which is
consistent with this interpretation.
When the Gauss constraint equation holds, however,
we find that the quantity $\fff\ffelec_{\hb\ib\jb}$ must be equal to
the tensor $\epsilon_{\hb\ib\jb}$ up to some constant $2C$.  Thus, we find
that $\fff(\hodge\fffield)=2C$.  The Lagrangian density is now written
$\DFFlgr=-2\volman\bigl(1/\fff\bigr)C^2$.  Thus, $C^2$ can be interpreted
as a (negative) cosmological constant when $\fff$ is a constant.  When
$\fff$ is not a constant, $\DFFlgr$ is essentially a contribution to the
dilaton potential proportional to $C^2$ times the
reciprocal of the coupling of the four-form field strength with the dilaton.
Although $C$ is a constant (rather than a scalar field), it is a constant
of integration rather than an imposed theoretical constant.

When the Gauss constraint holds, we can obtain an explicit expression for
the charge $\DFFcharge[\epsilon]$.  We see that
$\SFFCD^{ab}=6C\volbndini\epsilon^{ab}$, so the charge
$\DFFcharge[\epsilon]$
is just $12C$ times the area of the surface $\partial\ini$ whose volume
element is $\epsilon_{ab}$.


\sect{Dilaton Three-Form Sector}

In low energy string theory, one encounters a term in the action that is
a three-form field strength squared with a coupling to a dilaton.  Here,
we analyze such a gauge field with a somewhat more general coupling.
The three form field strength will be written as
$\tffield_{\lambda\mu\nu}=3!\,\del_{[\lambda}\tfpot_{\mu\nu]}$ where
$\tfpot_{\mu\nu}$ is a two-form potential field.  As always, there are
gauge transformations of this potential,
$\tfpot_{\mu\nu}[\chi]=\tfpot_{\mu\nu}[0]+2!\,\del_{[\mu}\chi_{\nu]}$,
under which the three-form field strength is invariant.
As usual, we can use this gauge freedom to guarantee that the components
of the potential in the orthonormal frame are finite everywhere in $\cal M$.
It is useful
to decompose the three-form field strength into `electric' and `magnetic'
pieces on a space-like hypersurface,
$\tfelec_{\ib\jb}=\projini^\lambda_\ib\projini^\mu_\jb\tffield_{\lambda\mu\nu}
\normini^\nu$ and
$\tfmagn=-(3!)^{-1}\epsilon^{\lambda\mu\nu\alpha}\tffield_{\lambda\mu\nu}
\normini_\alpha$ respectively.  It can then be shown that
$\tffield^{\lambda\mu\nu}\tffield_{\lambda\mu\nu}=3\tfmagn^2-6\tfelec^{\ib\jb}
\tfelec_{\ib\jb}$.

\subsection{The Dilaton-Three-Form Action}

As usual, the dilaton three-form action will be taken as the three-form
field strength squared coupled to some function of the dilaton.  Let this
function be given by $\ftf$, and make the usual assumptions about its form,
i.e., that it contain only the dilaton, but no derivatives of the dilaton.
Explicitly, we write the action:
\begin{equation}
  \DTFaction=\intman\volman{1\over2\cdot3!}\,\ftf\tffield^{\lambda\mu\nu}
  \tffield_{\lambda\mu\nu}.
  \label{DTF action}
\end{equation}
If we were to perform a duality rotation of the three-form field strength,
we would be left with a kinetic energy for a scalar field.  However, we are
not interested in this case here.  (In a sense, we have already considered
a scalar field in the context of the dilaton in which there is no coupling
to the Ricci scalar, though we have not considered a dilaton coupled to a
scalar field.)

The action of equation (\ref{DTF action}) can be subjected to variations in
the geometry, the dilaton, and the two-form potential.  The response in the
action is given by:
\begin{eqnarray}
  \var\DTFaction&=&\intman\volman\bigl(-\half\TFstress_{\mu\nu}
  \var\metric^{\mu\nu}-\half\TFdilsrc\var\dil+\EOMTF^{\mu\nu}\tfpot_{\mu\nu}
  \bigl)\nonumber\\
  & &\qquad+\intbnd\PiTF^{ij}\tfpot_{ij}-\intini\PTF^{\ib\jb}\tfpot_{\ib\jb}.
  \label{var DTF action}
\end{eqnarray}
As usual, the projection onto the boundary elements of the variation of the
potential are the same as the variation of the projection of the potential
onto the boundary elements.
When the variation of the two-form potential on the boundary is held fixed,
there is a contribution to the stress-energy-momentum tensor:
\begin{equation}
  \TFstress_{\mu\nu}=\ftf\bigl((2\cdot3!)^{-1}\metric_{\mu\nu}
  \tffield^{\alpha\beta\gamma}\tffield_{\alpha\beta\gamma}-\half
  \tffield_\mu{}^{\alpha\beta}\tffield_{\nu\alpha\beta}\bigr)
  \label{TF stress}
\end{equation}
as well as a dilaton source:
\begin{equation}
  \TFdilsrc=-{1\over6}\,{d\ftf\over d\dil}\,\tffield^{\lambda\mu\nu}
  \tffield_{\lambda\mu\nu}.
  \label{TF dil src}
\end{equation}
In addition, the equation of motion for the three-form field, in the
absence of a source, is the vanishing of
\begin{equation}
  \EOMTF^{\mu\nu}=-\del_\lambda\bigl(\ftf\tffield^{\lambda\mu\nu}\bigr).
  \label{EOM TF}
\end{equation}
The momenta conjugate to the two-form potential configuration on the boundaries
$\bnd$ and $\ini$ are, respectively,
\begin{equation}
  \PiTF^{ij}=\volbnd\ftf\normbnd_\lambda\tffield^{\lambda\mu\nu}\projbnd^i_\mu
  \projbnd^j_\nu
  \label{bnd TF momentum}
\end{equation}
and
\begin{equation}
  \PTF^{\ib\jb}=-\volini\ftf\normini_\lambda\tffield^{\lambda\mu\nu}
  \projini^\ib_\mu\projini^\jb_\nu=-\volini\ftf\tfelec^{\ib\jb}.
  \label{ini TF momentum}
\end{equation}

The potential $\tfpot_{ij}$ on the boundary $\bnd$ can be written in terms
of the quantities $V_a=\projbndini^i_a\tfpot_{ij}\normini^j$ and
$W_{ab}=\projbndini^i_a\projbndini^j_b\tfpot_{ij}$ on the
two-surface $\partial\ini$ as
\begin{equation}
  \tfpot_{ij}=2\normini_{[i}\projbndini_{j]}^a V_a+\projbndini_{[i}^a
  \projbndini_{j]}^b W_{ab}.
  \label{decomp tf pot}
\end{equation}
Note that since $W_{ab}$ is a two-form defined on the two-surface
$\partial\ini$, we could express it as $W_{ab}=w\epsilon_{ab}$ where
$w$ is a scalar function on the two surface.  Similarly, the variations
of $\tfpot_{ij}$ can be decomposed:
\begin{equation}
  \var\tfpot_{\ib\jb}=2\normini_{[i}\projbndini_{j]}^a\lapse^{-1}
  \bigl(\var(V_a\lapse)
  -W_{ab}\var\shift^b\bigr)+\projbndini_{[i}^a\projbndini_{j]}^b\var W_{ab}.
  \label{decomp var tf pot}
\end{equation}
This leads to a like decomposition of the momentum conjugate to the $\bnd$
boundary two-form potential.  We construct a surface three-form
charge density,
\begin{equation}
  \STFCD^a=2\volbndini\ftf\normbnd_\mu\tfelec^{\mu a}
  \label{stfcd}
\end{equation}
which is a vector on $\partial\ini$.  Note that $\STFCD^a$ is an extensive
quantity.  It can be used to define conserved charges as will be seen below.
Another extensive quantity is the surface three-form momentum density,
\begin{equation}
  \STFMD_b=\STFCD^a W_{ab}.
  \label{stfmd}
\end{equation}
Also, there is a surface three-form current density, defined as
\begin{equation}
  \STFCUR^{ab}=\volbndini\ftf\normbnd_\mu\tffield^{\mu ab}.
  \label{stfcur}
\end{equation}
Using these,
we can write the $\bnd$ boundary contribution to the variation of the
dilaton three-form action as:
\begin{equation}
  \var\DTFaction\vert_\bnd=\intbnd\bigl(\STFMD_a\var\shift^a
  -\STFCD^a\var(\lapse V_a)+\lapse\STFCUR^{ab}\var W_{ab}\bigr).
  \label{bnd var DTF action}
\end{equation}
Here, the first two terms in the integrand involve variations of intensive
variables with extensive coefficients while the last term is a variation
of an extensive variable with an intensive coefficient.

\subsection{Conserved Charges of the Three-Form Field}

Suppose that suitable matter is present such that there is a source term,
$\tfsrc^{\mu\nu}$, included in the equation of motion of the three-form
field.  Then we have from equation (\ref{EOM TF}),
\begin{equation}
  -\del_\lambda\bigl(\ftf\tffield^{\lambda\mu\nu}\bigr)=\tfsrc^{\mu\nu}.
  \label{EOM DTF src}
\end{equation}
However, we see from this equation that the source must be divergenceless,
$\del_\mu\tfsrc^{\mu\nu}=0$.  Thus, we can construct conserved charges
for it in the following way.  Let $f_\mu=\del_\mu\phi$ be an exact one-form.
($\phi$ is a scalar function.)  Due to the antisymmetry of $\tfsrc^{\mu\nu}$
in $\mu$ and $\nu$, we have
$0=f_\nu\del_\mu\tfsrc^{\mu\nu}=\del_\mu(\tfsrc^{\mu\nu}f_\nu)$.  If we
integrate this expression over $\cal M$, we have contributions from the
initial and final hypersurface as well as the boundary $\bnd$.  Supposing
that the source vanishes in the vicinity of $\bnd$, we find that the
quantity
\begin{equation}
  \DTFcharge[f]=-\intini\volini\normini_\mu\tfsrc^{\mu\nu}f_\nu
  \label{cons tf charge}
\end{equation}
is invariant upon the surface, $\ini$, of evaluation.  Thus it is a conserved
charge.

Using equation (\ref{EOM DTF src}), we can re-express the conserved charge
of equation (\ref{cons tf charge}) in terms of the two-surface three-form
charge density $\STFCD^a$.  Recall that $\normini_\mu=-\lapse\del_\mu\time$.
Then, the integrand of equation (\ref{cons tf charge}) can be written in
the form $\delini_\mu\bigl(\ftf\tfelec^{\mu a}f_a\bigr)$.  (Notice that
we have restricted the gradient $f_a$ to be a one-form on the dual tangent
space of $\partial\ini$.  Natural choices for the scalar $\phi$ are the
co\"ordinates of this hypersurface.)  Therefore, we have:
\begin{equation}
  \DTFcharge[f]=\intbndini\STFCD^a f_a
  \label{alt cons tf charge}
\end{equation}
where $f_a=\partial_a\phi$ and $\phi$ is an arbitrary scalar function on
$\partial\ini$.

\subsection{Canonical Form of the Dilaton-Three-Form Action}

The canonical decomposition of the action $\DTFaction$ is obtained through
the construction of the Hamiltonian density
$\DTFham=\PTF^{\ib\jb}\Lie_\time\tfpot_{\ib\jb}-\DTFlgr$.  A straightforward
calculation of the first term yields:
\begin{eqnarray}
  \PTF^{\ib\jb}\Lie_\time\tfpot_{\ib\jb}&=&\half\lapse\PTF^{\ib\jb}
  \tfelec_{\ib\jb}+\half\shift^\hb\tffield_{\hb\ib\jb}\PTF^{\ib\jb}\nonumber\\
  & &\qquad-2\tfpot_{\time\jb}\delini_\ib\PTF^{\ib\jb}+2\delini_\ib\bigl(
  \tfpot_{\time\jb}\PTF^{\ib\jb}\bigr)
  \label{DTF ham dens term 1]}
\end{eqnarray}
where $\tfpot_{\time\jb}=\time^\mu\tfpot_{\mu\jb}$ will act as a Lagrange
multiplier.  Also, the second term in the Hamiltonian density can be
written as:
\begin{equation}
  \DTFlgr=\lapse\volini\ftf(\fourth\tfmagn^2-\half\tfelec^{\ib\jb}
  \tfelec_{\ib\jb}).
  \label{DTF ham dens term 2}
\end{equation}
With these expressions, we can write the Hamiltonian:
\begin{eqnarray}
  \HDTF&=&\intini\DTFham\nonumber\\
  &=&\intini\bigl(\DTFhamcon\lapse+\DTFmomcon_\ib\shift^\ib+\tfpot_{\time\jb}
  \DTFgausscon^\jb\bigr)\nonumber\\
  & &\qquad+\intbndini\bigl(\STFCD^a\lapse V_a-\STFMD_a\shift^a\bigr).
  \label{DTF hamiltonian}
\end{eqnarray}
Here the contribution from the dilaton three-form sector to the Hamiltonian
constraint is given by:
\begin{equation}
  \DTFhamcon=-\fourth\volini\tfmagn^2
  \label{DTF ham con}
\end{equation}
the contribution to the momentum constraint is:
\begin{equation}
  \DTFmomcon_\hb=\half\tffield_{\hb\ib\jb}\PTF^{\ib\jb}
  \label{DTF mom con}
\end{equation}
and the Gauss constraint for the three-form field is:
\begin{equation}
  \DTFgausscon^\jb=2\delini_\ib\PTF^{\ib\jb}.
  \label{DTF gauss con}
\end{equation}
Also, the action of equation (\ref{DTF action}) can be re-expressed in the
canonical form.  It is:
\begin{eqnarray}
  \DTFaction&=&\intman\bigl(\PTF^{\ib\jb}\Lie_\time\tfpot_{\ib\jb}
  -\DTFlgr\bigr)\nonumber\\
  &=&\int dt\,\biggl(\intini\bigl(\PTF^{\ib\jb}\Lie_\time\tfpot_{\ib\jb}
  \nonumber\\
  & &\qquad\qquad
  -\DTFhamcon\lapse-\DTFmomcon_\ib\shift^\ib-\DTFgausscon^\ib\tfpot_{\time\ib}
  \bigr)\nonumber\\
  & &\qquad\qquad+\intbndini\bigl(\STFMD_a\shift^a-\STFCD^a\lapse V_a\bigr)
  \biggr).
  \label{canon DTF action}
\end{eqnarray}


\sect{Statistical mechanics}

Here we address, at a formal level, the construction of statistics of the
quantum mechanical theory based on the actions we have constructed.  The
mechanical theory is entirely described in terms of path integrals involving
our actions.  These yield quantum mechanical density matrices.  The
statistics of primary interest are the partition functions, which can
be thought of as functional integrals over the density matrices with
some periodic identification (with the period related to the temperature).

\subsection{The General Action and Canonical Ensembles}

We will consider a general form of action that possesses all of the
features of the actions we have considered thus far.
Let $A_\mu{}^\sss{A}$ denote the gauge field potential, where the upper-case
Latin index labels the various gauge fields that may be present.
The field strength
associated with the potential is given by $F^{\mu\nu}{}_\sss{A}$, and the
gauge field action functional, $\GFaction$,
involves the square of the field strength along with
a possible coupling to the dilaton, $\fgf$.  Note that for the three and
four form field strength gauge fields, the field index will actually be
a set of space-time indices, while for the Yang-Mills fields, it is a
colour index.
The total action of the theory is given by
$\totalaction=\EHDaction+\GFaction+\REFaction$.  Variations of this
action yield the usual equation of motion terms as well as boundary terms.
We will primarily be concerned with the latter.  The $\bnd$ portion of the
variation of the total action will take the form:
\begin{eqnarray}
  \var\totalaction\vert_\bnd&=&\intbnd\bigl(-\SED\var\lapse
  +\SMDTOT_a\var\shift^a-\SCD_\sss{A}\var(\lapse V^\sss{A})\nonumber\\
  & &\qquad+\lapse(\SSD^{ab}\var\fffbndini_{ab}+\SDD\var\dil
  +\SCUR^a{}_\sss{A}\var W_a{}^\sss{A})\bigr).
  \label{bnd var tot action}
\end{eqnarray}
Here, $\SCD_\sss{A}$ are the surface charge densities of the gauge fields,
$\SCUR^a{}_\sss{A}$ are the surface currents of the gauge fields,
and $\SMDTOT_a=\SMD_a+\SCD_\sss{A}W_a{}^\sss{A}$ is the total surface momentum
density.  Also, $V^\sss{A}$ are the projections of the $\bnd$ boundary
potentials
of the gauge fields along $\normini^i$, while $W_a{}^\sss{A}$ are the
portions of these fields restricted to $\partial\ini$.  Note that all of the
gauge field actions we have considered can be expressed in this way.
The action can be written in canonical form as:
\begin{eqnarray}
  \totalaction&=&\int dt\,\biggl(\intini\bigl(\momini^{\ib\jb}\Lie_\time
  \fffini_{\ib\jb}+\Pdil\Lie_\time\dil+P^\ib{}_\sss{A}\Lie_\time
  A_\ib{}^\sss{A}\nonumber\\
  & &\qquad\quad-\TOThamcon\lapse-\TOTmomcon_\ib\shift^\ib
  -\gausscon_\sss{A}A_\time{}^\sss{A}\bigr)\nonumber\\
  & &\qquad+\intbndini\bigl(-\SED\lapse+\SMDTOT_a\shift^a-\SCD_\sss{A}\lapse
  V^\sss{A}\bigr)\biggr).
  \label{canon tot action}
\end{eqnarray}
Here, $A_\time{}^\sss{A}$ are the projections of the gauge potentials
along the time vector.  $\TOThamcon$ is the total
Hamiltonian constraint, $\TOTmomcon_a$ is the total momentum constraint,
and $\gausscon_\sss{A}$ are the Gauss constraints of the gauge fields.

The first three terms in the integrand of equation (\ref{bnd var tot action})
involve variation of an intensive variable with an extensive coefficient.
However, the remaining terms involve variations of extensive variables with
intensive coefficients.  Define the micro-canonical action to be
\begin{equation}
  \microaction=\totalaction+\intbndini\bigl(\SED\lapse-\SMDTOT_a\shift^a
  +\SCD_\sss{A}\lapse V^\sss{A}\bigr).
  \label{micro action}
\end{equation}
Since this action differs from the original by boundary terms alone, the
equations of motion are unaffected.  However, such a transformation changes
the $\bnd$ boundary component of the variation of the action.  We now have:
\begin{eqnarray}
  \var\microaction\vert_\bnd&=&\intbnd\lapse\bigl(\var\SED-\rot^a\var\SMDTOT_a
  +V^\sss{A}\var\SCD_\sss{A}\nonumber\\
  & &\qquad+\half\SSD^{ab}\var\fffbndini_{ab}
  +\SDD\var\dil+\SCUR^a{}_\sss{A}\var W_a{}^\sss{A}\bigr)
  \label{bnd var micro action}
\end{eqnarray}
where we define $\lapse\rot^a=\shift^a$.  We will interpret $\rot^a$ as the
angular velocity of observers co-moving with the foliation on $\bnd$.
Each term in the integrand of
equation (\ref{bnd var micro action}) is of the form of a variation of an
extensive
variable with an intensive coefficient.  We see that the boundary
conditions imposed in order to obtain the equations of motion are those
in which the extensive variables are held constant.  It is for this reason
that the term `micro-canonical' has been adopted.  Also, note that the
Hamiltonian for the micro-canonical action contains only the constraints
(i.e., there are no additional boundary terms); so the canonical form
of the micro-canonical action has no $\partial\ini$ integral.  The importance
of this observation is that the micro-canonical action vanishes when
stationarity of all the fields and the constraint equations are imposed.

Grand-canonical boundary conditions involve the fixation of all {\em intensive}
variables on the boundary.  In analogy with the procedure above, we define
a grand-canonical action by:
\begin{equation}
  \grandaction=\totalaction-\intbnd\lapse(\half\SSD^{ab}\fffbndini_{ab}
  +\SDD\dil+\SCUR^a{}_\sss{A} W_a{}^\sss{A}).
  \label{grand action}
\end{equation}
Under variations of this action, the $\bnd$ boundary component becomes:
\begin{eqnarray}
  \var\grandaction\vert_\bnd&=&-\intbnd\bigl(\SED\var\lapse
  -\SMDTOT_a\var(\lapse\rot^a)+\SCD_\sss{A}\var(\lapse V^\sss{A})\nonumber\\
  & &\qquad+\fffbndini_{ab}\var(\lapse\SSD^{ab}/2)
  +\dil\var(\lapse\SDD)+W_a{}^\sss{A}\var(\lapse\SCUR^a{}_\sss{A})\bigr).
  \label{bnd var grand action}
\end{eqnarray}
so it is indeed the intensive boundary variables that must be fixed in order to
obtain the equations of motion.  In canonical form, the $\bnd$ contribution
to the grand-canonical action is given by:
\begin{equation}
  \grandaction\vert_\bnd=-\intbnd\lapse\bigl(\SDD-\SMDTOT_a\rot^a+\SCD_\sss{A}
  V^\sss{A}+\half\tr(\SSD)+\SDD\dil+\SCUR^a{}_\sss{A}
  W_a{}^\sss{A}\bigr).
  \label{bnd canon grand action}
\end{equation}

It will be useful to have a covariant form of the micro-canonical action.
To obtain this, we recall the definition of the micro-canonical action,
equation (\ref{micro action}), as well as the initial covariant form of
the total action, equation (\ref{EHD Action}) plus matter terms.  Since
the difference between the micro-canonical action and the total action
amounts to just a difference on the boundary $\bnd$, we will just attend
to this.  Thus:
\begin{equation}
  \microaction\vert_\bnd=\intbnd\bigl(-2\volbnd\feh\tr(\sffbnd)
  -\lapse\SED+\shift^a\SMDTOT_a-\lapse V^\sss{A}\SCD_\sss{A}\bigr).
  \label{bnd micro action}
\end{equation}
To cast this in a covariant form, we employ the definitions of $\SED$,
$\SMDTOT_a$ and $\SCD_\sss{A}$, as well as the decomposition given
in equation (\ref{decomp sff bnd}).  Then equation (\ref{bnd micro action})
can be re-written in the covariant form:
\begin{eqnarray}
  \microaction\vert_\bnd&=&\intbnd\volbnd\bigl(\feh\time_\mu\sffbnd^{\mu\nu}
  \partial_\nu\time-2\normbnd^\mu\partial_\mu\feh\nonumber\\
  & &\qquad-\fgf(\normbnd_\mu
  F^{\mu\nu}{}_\sss{A}\partial_\nu\time)(\time^\mu A^\sss{A}_\mu)\bigr).
  \label{cov bnd micro action}
\end{eqnarray}

\subsection{`Euclidean' Notation}

Until now, the metrics we have considered have been of Lorentzian signature.
Such metrics can be re-expressed in a `Euclidean' notation by redefining the
(real) metric functions in terms of new (complex) ones.  Our prescription
is the following:  We first rewrite the volume elements as
$\evolman=\im\volman$ and $\evolbnd=\im\volbnd$ and redefine the
lapse and shift functions: $\elapse=\im\lapse$ and $\eshift^\ib=\im\shift^\ib$.
The former leads to a new unit normal $\enormini_\mu=-\im\normini_\mu$ which
satisfies $\enormini\cdot\enormini=+1$.  In addition to the lapse and the
shift, we adopt a new notation for all the Lagrange multipliers in the
Hamiltonian.  Thus, ${\bar A}_\time{}^\sss{A}=\im A_\time{}^\sss{A}$.
Thus, the Hamiltonian as a functional of the Lagrange multipliers is
$\Bbb H[\elapse,\eshift^\ib,{\bar A}_\time{}^\sss{A}]=\im\Bbb H[\lapse,
\shift^\ib,A_\time{}^\sss{A}]$.  Finally, we define a new action functional
$I[\metric,\dil,A]=-\im S[\metric,\dil,A]$ such that the phase
in the path integral is written $\exp(-I)$.

An important feature of this prescription is the following.  Suppose that
a complex metric $\metric_\sss{\mathrm{C}}$ is obtained from a real, Lorentzian
metric $\metric_\sss{\mathrm{L}}$ via the transformation $\time\to-\im\time$.
Then, the Lagrange multipliers, $\lapse$, $\shift^\ib$, and $A_\time{}^\sss{A}$
become imaginary, or, equivalently, $\elapse$, $\eshift^\ib$, and
${\bar A}_\time{}^\sss{A}$ become real.  However, the canonical data,
that is $\momini^{\ib\jb}$, $\fffini_{\ib\jb}$, $\Pdil$, $\dil$,
$P^\ib{}_\sss{A}$ and $A_\ib{}^\sss{A}$ all are invariant.  Therefore,
{\em the extensive variables are all invariant under the Wick rotation\/}
since these variables are constructed out of the canonical data.
In particular, the values of the extensive variables of the complex metric
that extremize the path integral are the same as the values of these
variables on the corresponding Lorentzian metric.

We can write the micro-canonical and the grand-canonical actions using
the Euclidean notation in the canonical form.  They are:
\begin{eqnarray}
  \microeaction&=&\int dt\,\intini\bigl(-\im\momini^{\ib\jb}\Lie_\time
  \fffini_{\ib\jb}-\im P^\ib{}_\sss{A}\Lie_\time
  A_\ib{}^\sss{A}-\im\Pdil\Lie_\time\dil\nonumber\\
  & &\qquad\qquad+\TOThamcon\elapse+\TOTmomcon_\ib\eshift^\ib+\gausscon_\sss{A}
  {\bar A}_\time{}^\sss{A}\bigr)
  \label{canon eucl micro action}
\end{eqnarray}
and
\begin{eqnarray}
  \grandeaction&=&\microeaction+\int dt\,\intbndini\elapse\bigl(\SED
  -\SMDTOT_a\rot^a+\SCD_\sss{A}V^\sss{A}+\nonumber\\
  & &\qquad\qquad\half\tr(\SSD)+\SDD\dil+\SCUR^a{}_\sss{A} W_a{}^\sss{A}\bigl).
  \label{canon eucl grand action}
\end{eqnarray}
In addition, we can write out the variations of these actions.  For example,
the $\bnd$ boundary contribution of the variation of the Euclidean form of
the micro-canonical action can be constructed from equation
(\ref{bnd var micro action}).  It is:
\begin{eqnarray}
  \var\microeaction\vert_\bnd&=&-\int dt\,\intbndini\elapse\bigl(\var\SED
  -\rot^a\var\SMDTOT_a+V^\sss{A}\var\SCD_\sss{A}\nonumber\\
  & &\qquad\qquad+\half\SSD^{ab}\var\fffbndini_{ab}
  +\SDD\var\dil+\SCUR^a_\sss{A}\var W^\sss{A}_a\bigr).
  \label{bnd var micro eucl action}
\end{eqnarray}

\subsection{Functional Integrals}

It is well known that the classical equations of motion are obtained from
an action $\eaction$, by setting variations of this action to zero.  As
argued earlier, the variations are of a restricted class which fix certain
quantities on the boundary, and the quantities that must be fixed are
determined
by the choice of ensemble.  For example, it is the extensive boundary variables
that must be fixed in obtaining the equations of motion from the
micro-canonical
action.

This method of obtaining classical equations of motion is best understood
as a (classical) limit of the quantum mechanical density matrix obtained
from a path integral.  Formally, we can write this density matrix as:
\begin{equation}
  \dens=\int d[\metric,\dil,A]\,\e^{-\eaction[\metric,\dil,A]}
  \label{density matrix}
\end{equation}
where the integral is taken over all possible field configurations of the
metric, the dilaton, and the matter fields, from an initial and a final
space-like surface.  The field configurations are fixed on the initial and
the final space-like surfaces, $\initial$ and $\final$; so the density
matrix is a functional of these quantities.  In addition there will be
quantities on $\bnd$ joining the boundaries of these two space-like surfaces
that must also be fixed.  These quantities depend on the ensemble chosen
for the action.  The density matrix will be a functional of these variables
as well.  The density matrix gives us all the quantum mechanical information
we need to solve most problems.  Furthermore, the classical limit can be
understood in the following way.  Given an initial configuration, there
is a final configuration that is most `likely' in a quantum-mechanical
sense: it is the one for which the density matrix has the largest value.
In the classical limit, the action is always much larger than unity
(in units of Planck's constant).  Thus, the final configuration that
extremizes the density matrix is the one for which $\var\eaction=0$, that
is, the one for which the classical field equations hold.  (Note that
$\eaction$ is a complex quantity---we have not yet made a Wick
rotation---so the last statement follows from stationary phase arguments.)

Much information about the system can be obtained by taking statistics
of the density matrix.  The most important of these is known as the
partition function.  It is obtained by `tracing over' the initial and
final states.  In terms of path integrals, this can be realized by
identifying the initial and the final configurations, and integrating
over all possible configurations.  The partition function then depends
upon the information fixed on the boundary $\bnd$ alone.  Note that,
in identifying the initial and the final space-like surfaces, we have
effectively changed the topology of $\cal M$ from $\ini\times{\cal I}$
to $\ini\times S^1$.

As an example, consider the micro-canonical ensemble.  The objects that
are held fixed when this action is varied are the metrics and field
configurations on the initial and final surfaces as well as the
extensive variables on the boundary $\bnd$.
Thus, the functional dependence of the density matrix is:
\begin{eqnarray}
  \dens&=&\dens[\fffini_{\mathrm{i}},\dil_{\mathrm{i}},A_{\mathrm{i}};
  \fffini_{\mathrm{f}},\dil_{\mathrm{f}},A_{\mathrm{f}};
  \SED,\SMDTOT,\fffbndini,\ldil,\SCD,W]\nonumber\\
  &=&\int d[\metric,\dil,A]\,\e^{-\microeaction[\metric,\dil,A]}.
  \label{micro dens}
\end{eqnarray}
The subscripts `i' and `f' refer to the initial and final space-like
hypersurfaces, $\initial$ and $\final$, respectively and $\ldil$ is the
dilaton configuration on $\bnd$.
The partition function is also known as the density of states.
It is given by:
\begin{equation}
  \nu[\SED,\SMDTOT,\fffbndini,\ldil,\SCD,W]
  =\int d[\fffini,\dil_\ini,A_\ini]\,\dens[\fffini,\dil_\ini,A_\ini;
  \fffini,\dil_\ini,A_\ini;\SED,\SMDTOT,\fffbndini,\ldil,\SCD,W].
  \label{micro part fn}
\end{equation}
Note that we have explicitly indicated the periodic identification of the
fields on the initial and final space-like boundaries in the density
matrix.

In a similar manner, we can construct a grand-canonical density matrix
by making a Laplace transform of equation (\ref{micro dens}).  This
density matrix would be a functional of $\rectemp$, $\rot^a$, $\SSD_{ab}$,
$\SDD$, $V^\sss{A}$, and $\SCUR_\sss{A}$ on the boundary $\bnd$.  Here,
the reciprocal temperature, $\rectemp$, is defined by:
\begin{equation}
  \rectemp=\oint dt\,\elapse\vert_{\partial\ini}
  \label{rec temp}
\end{equation}
where we have used $\oint$ as a reminder that the initial and final space-like
hypersurfaces have been periodically identified; $\rectemp$ is the gauge
invariant measure of this period.
The grand-canonical partition function is then
${\cal Z}[\rectemp,\rot,\SSD,\SDD,V,\SCUR]$.


\sect{Black Hole Thermodynamics}

The thermodynamics of systems containing black holes can be obtained
using the statistical techniques of the previous section.  In particular,
we wish to obtain a thermodynamic `first law' which relates the entropy
change of a system with changes in extensive quantities of the system.
Clearly, such a result should be obtained from the micro-canonical action.
The entropy is normally defined as the logarithm of the density of states.
(We recall that the latter is just the partition function for the
micro-canonical ensemble.)  We will only consider black hole solutions that
are stationary in the sense that the Lie derivative of all fields (including
the metric, dilaton, and matter fields) vanish.  Also, we continue to use
the `generalized gauge fields' introduced in the previous section.  Any
of the gauge fields discussed in this paper will have a similar form.

We will only construct the `zeroth order' contribution to the entropy
from the path integral for the density of states.  In this approximation,
the density of states is written:
\begin{equation}
  \nu[\SED,\SMDTOT,\fffbndini,\ldil,\SCD,W]\approx\e^{-\microeaction}
  \label{zero order micro part fn}
\end{equation}
where the micro-canonical action is evaluated at its (complex)
extremum value.  Clearly, the entropy is just:
\begin{equation}
  \entropy[\SED,\SMDTOT,\fffbndini,\ldil,\SCD,W]\approx-\microeaction
  \vert_\sss{\mathrm{CL}}.
  \label{entropy}
\end{equation}
Thus, to evaluate the entropy is to evaluate the micro-canonical action
at its (complex) extremum value.  However, the presence of an event
horizon introduces some interpretational problems.  These we address in
the rest of this section.

\subsection{Regularizations at the Event Horizon}

As usual, we turn to the Euclidean form of the action.  The system will
be defined as the interior to some closed two surface, $\bouter$, on which
the micro-canonical boundary data are specified.  This interior will include
an event horizon; the foliation will become degenerate on this horizon.
In addition to the usual foliation in the time-like direction, we will
suppose that it is possible to foliate from the event horizon to the outer
boundary.  After periodic identification, the complex manifold has the
form of a cone$\times S^2$, where the azimuthal direction of the cone
is the time-like direction and the `radial' direction corresponds to the
space-like vector along which the space-like foliation was constructed.
The outer edge of the cone is the two-surface $\bouter$, and the `point'
of the cone is where the time-like foliation becomes degenerate: the
event horizon.

It is common to impose regularity conditions on the cone to cause it to
become a disk.  Near the event horizon, we {\em assume\/} that it is possible
to write the metric in the Euclidean form:
\begin{equation}
  ds^2_\sss{\mathrm{E}}\simeq\elapse^2 d\time^2+\elevate^2 d\radius^2
  +\fffbndini_{ab}dx^a dx^b.
  \label{approx e metric}
\end{equation}
Here, $\radius$ is the co\"ordinate of foliation in the space-like
direction, and $\elevate$ is the analog of the lapse function defined
as $\normbnd_\mu=\elevate\partial_\mu\radius$.  The conical singularity
is not present in the metric of equation (\ref{approx e metric}) if the
`circumference' of the circles of constant $\time$ have the value of
$2\pi$ times the proper radius near the event horizon.  Let $\period$ be the
period of the identification of the co\"ordinate $\time$.  Then the
circumference of the circles of constant $\time$ is given by $\elapse\period$,
and the regularity condition is $\period(\normbnd^\ib\partial_\ib\elapse)=2\pi$
at the event horizon.  Choose the period $\period$ to satisfy this condition.
Then it takes the value $\period=2\pi/\ehgrav$ where
$\ehgrav^2=\fffini^{\ib\jb}(\partial_\ib\elapse)(\partial_\jb\elapse)$
(evaluated on the event horizon) is the surface gravity of the event
horizon.

\subsection{Evaluation of the Micro-Canonical Action}

Although the event horizon is in no way distinctive to an infalling observer,
nevertheless, for a system observer (i.e., one who lives on the system
boundary $\bouter$) it represents a one-way membrane onto which information
can approach, but from which nothing can come (at least classically).
For such an observer, it would be inappropriate to attempt
to calculate an action that included the event horizon.  So let us remove
an open set surrounding the event horizon infinitesimally close.  The
remaining portion of the manifold now contains a new `boundary.'  However,
this is {\em not} a system boundary on which thermodynamic data must be
specified---it is just a tool that will represent the information lost
in `throwing away' the event horizon from the system.  With this understanding,
we shall denote this surface as $\binner$.

Let us recall the covariant form of the micro-canonical action.  It is
the usual action but with the $\bndout$ boundary term given by
equation (\ref{cov bnd micro action}).  There is no supplemented boundary
term for the inner boundary.  Using the usual techniques to perform a
canonical decomposition of this action, and keeping track of the terms
appearing on the new boundary $\binner$, we find that
\begin{eqnarray}
  \asteaction=\microeaction&+&\int dt\,\int_
  {\hbox to 1em{$\scriptstyle\binner\hss$}}
  d^2\!x\bigl(2\volbndini\feh
  \normbnd^\ib\partial_\ib\elapse-2\volbndini\elapse\normbnd^\ib\partial_\ib
  \feh\nonumber\\
  & &\qquad\qquad-\eshift^a\SMDTOT_a+\elapse\SCD_\sss{A}V^\sss{A}\bigr).
  \label{canon eucl ast action}
\end{eqnarray}
in the `Euclidean' notation.
Here, $\asteaction$ is the thermodynamic action which differs from the
micro-canonical action, $\microeaction$, in that the event horizon
has been removed from the micro-canonical system.
In obtaining equation (\ref{canon eucl ast action}), we have used the
relation $a_\ib=\elapse^{-1}\partial_\ib\elapse$.
The action $\microeaction$ was given in equation
(\ref{canon eucl micro action}).
Thus, we notice that the only boundary contribution to $\asteaction$ is
on the boundary $\binner$.

We wish to evaluate the action $\asteaction$ on the solution in which the
field equations hold---the extremal solution.  The contribution from
the action $\microeaction$ of equation (\ref{canon eucl micro action})
vanishes:
the terms involving Lie derivatives of the
fields vanish due to the stationarity
of the solution, and the remaining terms vanish because the constraint
equations must hold on a classical solution.  Thus the sole contribution
to the action $\asteaction$ comes from the boundary contribution $\binner$
of equation (\ref{canon eucl ast action}) which arises due to the discarding
of the event horizon from the system.  In evaluating this term, we invoke
the regularity conditions of the previous sub-section: $\elapse=0$,
$\eshift^a=0$, and $\int\normbnd^\ib\partial_\ib\elapse dt=-2\pi$.
(The shift approaches zero as we approach the event horizon since we have
chosen a zero-vorticity observer.  The negative sign in the third equation
arises since the normal vector is now {\em inward\/} directed, rather than
outward as it was in the previous sub-section.)  Thus we find that:
\begin{equation}
  \asteaction\vert_\sss{\mathrm{CL}}=-4\pi\intbndini\volbndini\feh.
  \label{cl eucl ast action}
\end{equation}
The entropy (neglecting higher-order quantum corrections) is just
the negative of this quantity according to equation (\ref{entropy})
(using $\asteaction$ rather than $\microeaction$).  The usual value
for the entropy with no dilaton field---one quarter of the event
horizon area---is recovered when we set $\feh=(16\pi)^{-1}$. (This is the
appropriate value for this coefficient for the usual Einstein-Hilbert
action in units where Newton's constant is unity.)

\subsection{The First Law of Thermodynamics}

Since the inner boundary, $\binner$, is not really a boundary, we will
view $\asteaction$ as a functional of the same extensive variables
on $\bouter$ as $\microeaction$.  The first law of thermodynamics is
obtained by varying the entropy given by equation (\ref{entropy}) with
the aid of equation (\ref{bnd var micro eucl action}) which applies
equally to $\asteaction$.  These variations are understood to be amongst
those that preserve the classical equations of motion, so the boundary
contribution is the only one present.  (Recall that we have closed the
manifold with respect to the initial and final hypersurfaces, so there
are no longer any initial and final space-like boundaries.)  Thus:
\begin{equation}
  \var\entropy=\int_{\hbox to 1em{$\scriptstyle\bouter$\hss}}
  d^2\!x\,\rectemp\bigl(\var\SED-\rot^a\var\SMDTOT_a
  +V^\sss{A}\var\SCD_\sss{A}+\half\SSD^{ab}\var\fffbndini_{ab}
  +\SDD\var\dil+\SCUR^a{}_\sss{A}\var W_a{}^\sss{A}\bigl)
  \label{first law}
\end{equation}
is our formulation of the first law of thermodynamics.  Recall that
$\rectemp$ is the reciprocal temperature of the system
(equation (\ref{rec temp})). Note that it is not necessarily constant over
the system boundary, so we have left it  within the integral.  When the
boundary is chosen to be an isothermal surface, then we can integrate
the first term in the integrand of equation (\ref{first law}) to obtain
the usual `$\rectemp\var\energy$' term.  (Recall that $\energy$ is the
quasilocal energy of equation (\ref{quasilocal energy}).)  However, the
angular velocity $\rot^a$ will can not generally be a constant on the system
boundary simultaneously, so the usual expression for the first law does
not generally hold in the case of a finite system. In the case of
non-abelian fields, if a surface can be chosen such that $\rectemp
V^\sss{A}$ is constant and proportional to the gauge Killing scalar
(if such exists), then the contribution of this term in the first
law is of the form $\rectemp V \var\DYMcharge$ where the
latter quantity is defined in (\ref{conserved YM charge}).
We note that the formulation (\ref{first law})
of the first law differs from that considered previously for
non-abelian gauge fields \cite{Sud}, in which certain asymptotic
properties of the gauge and gravitational fields were assumed in order
to define a colour charge---the resultant formulation of the
first law is therefore valid only in this asymptotic region.
Our formulation (\ref{first law}) of the first law recovers  this result
when the gauge and gravitational fields have the aforementioned
falloff properties.

It is useful to divide the variations of the metric
on $\bouter$ into a `shape' preserving piece and a `volume'
(which is, of course, an area on a two-surface) preserving
piece as follows:
$\var\fffbndini_{ab}=\shapebndini_{ab}\var\volbndini
+\volbndini\var\shapebndini_{ab}$,
where $\shapebndini_{ab}=\fffbndini_{ab}/\volbndini$.
If this is done, then the $\half\rectemp\SSD^{ab}\var\fffbndini_{ab}$ term in
the integrand of equation (\ref{first law}) can be re-written as
$\rectemp(\spress\var\VOL+\sshape^{ab}\var\shapebndini_{ab})$.  Here,
$\VOL=\volbndini$ can be thought of as a measure of the volume
of the system (by which we really mean the area of $\bouter$), and
thus $\spress=\half\tr(\SSD)/\volbndini$ is interpreted as the
pressure on the system.  The quantity
$\sshape^{ab}=\half\volbndini\SSD^{ab}$
is thermodynamically conjugate to shape changes of the system boundary.


\sect{Concluding Remarks}

The formulation of the first law of thermodynamics as given in
(\ref{first law}) generalizes previous formulations \cite{Hayward,BMY} to
include the most general couplings of gauge fields to dilatonic
gravity  in four
spacetime dimensions that have at most two derivatives in any term in
the action.  We close by commenting on possible extensions of our
work.

General arguments from string theory suggest that the action
considered in this paper receives corrections from terms that have
more than two derivatives in the metric and matter fields, i.e.,
terms which are at least quadratic in the curvature and/or field
strengths. Although the general form for the Noether charge for such terms
has been evaluated \cite{IW}, the detailed manner in which such terms
contribute to the first law of thermodynamics in the context of the
quasilocal formalism considered in this paper remains to be worked
out.

A related problem of interest concerns the r\^ole of topological fields.
These are fields which couple to the connection but not to the
metric. It has recently been shown that interesting black hole
solutions exist for a model topological field theory in $(2+1)$
dimensions \cite{Cargeg}. Their thermodynamical properties are
considerably different from the usual case \cite{BCM,BTZ} and are not
fully understood.  The generalization of the quasilocal formalism to
such actions represents an interesting problem since the first
derivatives of the metric will play a markedly different r\^ole in the
boundary terms.

\paragraph*{Acknowledgements:}
We are grateful for the hospitality
of D.A.M.T.P., where this work was carried out, and for financial
support by the Natural
Sciences and Engineering Research Council (NSERC) of Canada.

\vfill\pagebreak


\appendix
\sect{Notation}

This is a summary of the notation used in this paper.
The manifold $\cal M$ is topologically $\ini\times{\cal I}$ where
$\ini$ is a spacelike hypersurface and $\cal I$ is a timelike interval.
Foliation of $\cal M$ along $\cal I$ allows us to define the
lapse, $N$, and the shift, $N^\mu$.  The various manifolds we will
consider, and some of the tensors defined on them are summarized in
Table~1.
\begin{table}[h]
\caption{Manifold Variables}\vspace{3pt}
\begin{center}
\begin{tabular}{lcccc}
  \hline
  &\multicolumn{4}{c}{Manifold}\\
  \cline{2-5}
  & $\cal M$ & $\bnd$ & $\ini$ & $\partial\ini$ \\
  \hline\hline
  Indicies & $\{\mu,\nu,\ldots\}$ & $\{i,j,k\}$ & $\{\hb,\ib,\jb\}$
    & $\{a,b\}$ \\
  Normal Vector & -- & $\normbnd_\mu$ & $\normini_\mu$ & $\normini_i$ \\
  Metric & $\metric_{\mu\nu}$ & $\fffbnd_{ij}$ & $\fffini_{\ib\jb}$
    & $\fffbndini_{ab}$ \\
  Compatible Derivative & $\del_\mu$ & $\delbnd_i$ & $\delini_\ib$ & -- \\
  Intrinsic Curvature Scalar & $\Ricci[\metric]$ & -- & $\Ricci[\fffini]$
    & -- \\
  Extrinsic Curvature & -- & $\sffbnd_{ij}$ & $\sffini_{\ib\jb}$
    & $\sffbndini_{ab}$ \\
  Geometric Momentum & -- & $\mombnd^{ij}$ & $\momini^{\ib\jb}$ & -- \\
  \hline
\end{tabular}
\end{center}
\end{table}
We can construct the following densities on $\partial\ini$ out of projections
of $\mombnd^{ij}$: the surface energy density, $\SED$, the surface geometric
momentum density, $\SMD_a$, the surface stress density $\SSD^{ab}$.

We consider various forms of matter.  These are summarized in Tables~2 and~3.
Table~2 contains the field variables used in the various sectors.
\begin{table}[h]
\caption{Field Variables}\vspace{3pt}
\begin{center}
\begin{tabular}{lcccccc}
  \hline
  &  \multicolumn{3}{c}{Field Strengths} & \multicolumn{3}{c}{Potential Fields}
    \\
  \cline{2-4} \cline{5-7}
  & full & electric & magnetic & full & along $\normini$ & on $\partial\ini$
    \\
  \hline\hline
  Dilaton & \multicolumn{3}{c}{-- $\ddil$ --}
    & \multicolumn{3}{c}{-- $\dil$ --} \\
  Axion & \multicolumn{3}{c}{-- $\daxion$ --}
    & \multicolumn{3}{c}{-- $\axion$ --} \\
  Yang-Mills & $\Faraday_{\mu\nu}{}^\fa$ & $\elec_i{}^\fa$ & $\magn_i{}^\fa$
    & $\conn_\mu{}^\fa$ & $\fV^\fa$ & $\fW_a{}^\fa$ \\
  Four-Form & $\fffield_{\mu\nu\rho\sigma}$ & $\ffelec_{\hb\ib\jb}$ & --
    & $\ffpot_{\lambda\mu\nu}$ & $V_{ab}$ & -- \\
  Three-Form & $\tffield_{\lambda\mu\nu}$ & $\tfelec_{\ib\jb}$ & $\tfmagn$
    & $\tfpot_{\mu\nu}$ & $V_a$ & $W_{ab}$ \\
  General Field & $F_{\mu\nu}{}^\sss{A}$& & & $A_\mu{}^\sss{A}$ & $V^\sss{A}$
    & $W_a{}^\sss{A}$ \\
  \hline
\end{tabular}
\end{center}
\end{table}
The momenta corresponding to these fields (and the related densities
constructed on $\partial\ini$) are given in Table~3.
\begin{table}[h]
\caption{Conjugate Momenta}\vspace{3pt}
\begin{center}
\begin{tabular}{lccccc}
  \hline
  & \multicolumn{2}{c}{Conjugate Momenta} & \multicolumn{3}{c}{Surface Density}
    \\
  \cline{2-3}\cline{4-6}
  & on $\ini$ & on $\bnd$ & charge & momentum & current \\
  \hline\hline
  Dilaton & $\Pdil$ & $\Pidil$ & \multicolumn{3}{c}{-- $\SDD$ --} \\
  Axion & $\Paxion$ & $\Piaxion$ & \multicolumn{3}{c}{-- $\SAD$ --} \\
  Dilaton-Yang-Mills & $\PYM^\ib{}_\fa$ & $\PiYM^i{}_\fa$ & $\SYMCD_\fa$
    & $\SYMMD_a$ & $\SYMCUR^a{}_\fa$ \\
  Axion-Yang-Mills & $\PAYM^\ib{}_\fa$ & $\PiAYM^i{}_\fa$ & $\SAYMCD_\fa$
    & $\SAYMMD_a$ & $\SAYMCUR^a{}_\fa$ \\
  Dilaton-Four-Form & $\PFF^{\hb\ib\jb}$ & $\PiFF^{ijk}$ & $\SFFCD^{ab}$
    & -- & -- \\
  Dilaton-Three-Form & $\PTF^{\ib\jb}$ & $\PiTF^{ij}$ & $\STFCD^a$ & $\STFMD_a$
    & $\STFCUR^{ab}$ \\
  General Field & $P^\ib{}_\sss{A}$ & & $\SCD_\sss{A}$ & & $\SCUR^a{}_\sss{A}$
    \\
  \hline
\end{tabular}
\end{center}
\end{table}

\vfill
\clearpage


\end{document}